\documentclass[journal]{IEEEtran}
\usepackage{caption}

\usepackage{tikz}
\usepackage{graphicx}

\usepackage{amsmath}

\usepackage{amssymb}

\usepackage{tikz-3dplot}
\usepackage{comment}
\usepackage{cite}
\usepackage{caption}
\usepackage{subcaption}

\usepackage{scrextend}
\usepackage{algorithm}
\usepackage{algpseudocode}

\ifCLASSINFOpdf
\else

\fi
\usepackage[switch]{lineno}
\usepackage{setspace}
\begin{document}

\title{Enhancing Resilience Against Jamming Attacks: A Cooperative Anti-Jamming Method Using Direction Estimation}

\author{Amir Mehrabian and  Georges Kaddoum,~\IEEEmembership{Senior Member,~IEEE}

\thanks{Amir Mehrabian and Georges Kaddoum are with the LaCIME Laboratory,
Department of Electrical Engineering, École de Technologie Supérieure,
Montreal, QC H3C 0J9, Canada (e-mail: \{amir.mehrabian, georges.kaddoum\}@etsmtl.ca).}}
\markboth{}%
{Shell \MakeLowercase{\textit{et al.}}: Bare Demo of IEEEtran.cls for IEEE Journals}

\maketitle

\begin{abstract}
The inherent vulnerability of wireless communication necessitates strategies to enhance its security, particularly in the face of jamming attacks. This paper uses the collaborations of multiple sensing nodes (SNs) in the wireless network to present a cooperative anti-jamming approach (CAJ) designed to neutralize the impact of jamming attacks. We propose an eigenvector (EV) method to estimate the direction of the channel vector from pilot symbols. Through our analysis, we demonstrate that with an adequate number of pilot symbols, the performance of the proposed EV method is comparable to the scenario where the perfect channel state information (CSI) is utilized. Both analytical formulas and simulations illustrate the excellent performance of the proposed EV-CAJ under strong jamming signals. Considering severe jamming, the proposed EV-CAJ method exhibits only a 0.7 dB degradation compared to the case without jamming especially when the number of SNs is significantly larger than the number of jamming nodes (JNs). Moreover, the extension of the proposed method can handle multiple jammers at the expense of degrees of freedom (DoF). We also investigate the method's ability to remain robust in fast-fading channels with different coherence times. Our proposed approach demonstrates good resilience, particularly when the ratio of the channel's coherence time to the time frame is small. This is especially important in the case of mobile jammers with large Doppler shifts.
\end{abstract}




\IEEEpeerreviewmaketitle

\section{Introduction}\label{intro_sec}
Wireless Sensor Networks (WSNs) are integral to a wide range of internet of things (IoT) applications, from environmental monitoring to smart cities and industrial automation. As the applications of WSNs continue to expand, the threats they face become increasingly concerning, highlighting the urgent need for effective security measures. Since these networks collect and transmit sensitive data, their security becomes critical. Thus, WSNs must be protected to ensure data integrity and maintain a reliable performance \cite{survey_jamming}. \par
As a wireless communication system, WSNs rely on open channels for data transmission, and the air interface serves as the medium for signal transmission. Consequently, the open nature of wireless channels makes them vulnerable to unauthorized entities, such as eavesdroppers, which can monitor signals transmitted through the wireless channel. Moreover, the wireless channel is unprotected against interference sources and malicious jammers. These malicious jamming nodes (JNs) can intentionally transmit strong signals to disrupt and block the communication among authorized entities within a wireless communication network \cite{fakoorian_survey}.
A simple, inexpensive transmitter can still carry out damaging jamming attacks \cite{survey_jamming}, while more sophisticated jamming methods, such as reactive and deceptive jammers, are even more efficient, stealthy, and devastating \cite{Krishna_fhss}.
\par
WSNs are further resilient against fading thanks to the presence of multiple sensing nodes (SNs), also referred to as degrees of freedom (DoF). The existence of multiple SNs and their ability to cooperate can be exploited to strengthen the robustness of communication systems, particularly against jamming attacks. In the literature, various anti-jamming (AJ) techniques have been proposed to improve the security of these systems, specifically against jamming attacks, such as avoiding jamming signals or suppressing their effects \cite{PLS_survey, Kaddoum_survey}.\par

\subsection{Related Works}
Numerous works study approaches at the physical layer to increase the security of communication systems \cite{Kaddoum_survey}. Physical layer security (PLS) techniques employ the inherent characteristics of wireless channels to provide an additional layer of protection.
PLS methods, including nullification and friendly jamming, have been extensively studied to enhance secrecy rates across various novel communication techniques and network configurations
\cite{PLS_new1,PLS_new2, PLS_new3, reviewer_cite}. 
In addition, a wide variety of approaches utilize physical layer techniques to enhance resilience against active jamming attacks. 
For instance,  spread spectrum methods, such as frequency hopping spread spectrum (FHSS) \cite{Fhss_popper}, are commonly regarded as an anti-jamming (AJ) strategy for hopping into non-jammed channels. However, intelligent and smart jammers can sense the spectrum and discover the employed channels to target active channels \cite{Krishna_fhss,survey_jamming}.
In addition, spread spectrum AJ methods, such as FHSS and  direct sequence spread spectrum (DSSS) \cite{DSSS_wlan},  are inefficient in terms of   spectrum utilization \cite{survey_jamming}. A highly powerful jamming signal targeting multiple channels cannot be effectively managed using FHSS techniques \cite{jam_me_if_you_can}. \par
Several anti-jamming strategies that use reinforcement learning (RL) and game theory to allocate resources have been recently developed to avoid jammed channels. Many of these techniques aim to evade jammers by employing frequency hopping to free channels, which requires a broader available spectrum for this purpose \cite{jam_me_if_you_can}. However, a sophisticated jammer could anticipate such tactics and target that selected channel,  this may lead to a Stackelberg game scenario  \cite{stackel_0,stackel_1,stackel2,Kaddoum_stackel,RL-FH}. The introduced method in \cite{stackel_0} models jammer-user interactions as a game, with users employing frequency hopping to evade jamming. A bimatrix game framework is used to derive the Nash Equilibrium (NE) strategy for improved performance. The authors of \cite{stackel_1} study cooperative AJ through a Stackelberg game framework for channel selection and power allocation, where users act as leaders and a single jammer follows. The work proposed a mechanism to sacrifice the benefits of a few users to trap the jammer, allowing the remaining users to achieve improved performance. In \cite{stackel2}, a game theory framework is also applied to a multi-user scenario, proposing two collision avoidance protocols to allow only one user to communicate for improved performance. In \cite{Kaddoum_stackel},  a non-cooperative Stackelberg game for multi-channel jamming is presented with an RL solution to determine power allocation for victim channels.
Typically, game-theory and RL methods require several vacant channels for frequency hopping to hide from the JN. Additionally,  game-theory methods usually require prior knowledge of jamming actions and complete information about channels, which is not practical \cite{survey_jamming}. Furthermore, the RL methods, particularly those based on Q-learning, face challenges when dealing with large state and action spaces. This often leads to slow convergence  and difficulties in learning the jammer's behavior \cite{jam_me_if_you_can}.\par
Furthermore, several methods in the literature use multiple-input multiple-output (MIMO) systems in order to combat smart jammers at the physical layer. In contrast to AJ approaches that attempt to avoid jamming, the MIMO-based methods try to retrieve the desired signal even under heavy jamming conditions. In urban environments, several AJ methods based on MIMO were proposed which use estimation of the jamming channel to nullify the jamming signal \cite{conf_hoang, Multiple_silent_antijam}. These works employ the embedded silent period in the time frame where only a jamming signal exists for the estimation of jamming channels. However, a smart detector capable of performing a reactive attack is able to rapidly stop its transmission in order to save its energy and avoid identification of its signal. In \cite{pilot_mMIMO,pilot_sequence_mMimo2,pilto_MIMO3,pilot_MIMO4}, massive MIMO systems were studied for jamming removal. These studies use the orthogonality of the pilot sequences and the large number of antennas in massive MIMO to estimate the complete channel state information (CSI) of the jamming channel. Consequently, the CSI of the jamming channel can be applied for the removal of the jamming signal. However, the CSI estimation process  is complicated due to the increased number of antennas. It is likely that the assumption of independent fading channels is violated because of correlated channels caused by closely spaced antenna elements \cite{Kaddoum_survey}. In addition, these studies assume a block fading model and do not investigate the effects of fast fading and channel variations within a block, which contradicts the proposed model.
Zero-forcing interference cancellation (ZFIC) is proposed in \cite{ZFIC3_conf,ZFIC}, where the proposed method estimates the jamming channel during a silent interval where the friendly transmitting node (TN) is inactive. Moreover,  \cite{ZFIC}  considers multiple sources of jammer and extends the ZFIC with majority voting (ZFIC-MV). Additionally, in the context of orthogonal frequency-division multiplexing (OFDM), a successive interference cancellation method for jamming removal is proposed. This method relies on estimating the friendly channel during the preamble interval, assuming the  JN is inactive. \cite{ZFIC-conf}. The authors of  \cite{MAED,MADE_conf}  consider the simultaneous optimization of the mitigation, estimation, and detection (MADE) for massive MIMO systems in the presence of only one JN. The suggested optimization problem is non-convex and is solved using the splitting step method. \par 
\subsection{Contributions}
In the literature,  a limited number of studies have used collaboration among wireless SNs to eliminate the signals of one or multiple malicious JNs. In this work, we consider a cooperative scenario, where multiple SNs can collaboratively perform effective jamming cancellation. Our proposed method does not require a silent interval in which the JN or the TN is silent for CSI estimation, in contrast to  \cite{conf_hoang, Multiple_silent_antijam}. We employ the pilot block for the removal of friendly signals, and then eigenvectors of the received signal matrix are used to estimate the JN's channel direction only. For multiple JNs, we assume that a source separation method  \cite{MDL,Vinogradova,Nayebi} can identify the number of active JNs.
By determining the number of active  JNs, the proposed method can apply a rank constraint. This allows the proposed method to estimate only the spanning vectors of the jamming channel space. When one jammer is active, the jamming channel space is linear and aligned with the direction of the jamming channel vector. Thus, our proposed method only estimates that specific direction instead of estimating the entire vector and CSI of the jammer, in contrast to \cite{pilot_mMIMO,pilot_sequence_mMimo2,pilto_MIMO3} and several game-theoretic AJ methods. Therefore, the proposed method can identify a matrix based on the complementary space of all the jamming channel vectors. This complementary space is orthogonal to all the jamming channel vectors, enabling not only the effective elimination of a single jamming signal but also the suppression of multiple jamming signals, in contrast to \cite{MAED,MADE_conf}, which can handle only one JN. As a result, the proposed method demonstrates an improving performance, closely approaching scenarios where no jamming is present, particularly when the number of SNs becomes much larger than the number of the JNs. In addition, this paper thoroughly investigates the effect of a fast-fading jamming channel on the proposed methods to fully evaluate their performance under realistic channel conditions.\par
Moreover, our work exploits the degrees of freedom (DoF) resulting from the cooperation of sensing nodes (SNs) to neutralize jamming effects. This method can be used alongside techniques, such as frequency hopping and channel selection based on game theory or RL, as described in \cite{stackel_1,stackel2,wang,Yadong}. In contrast to these works, given that our method focuses on retrieving the desirable signal from the jammed signal rather than avoiding it or hiding from it \cite{jam_me_if_you_can}, the proposed method remains applicable even in scenarios where all channels are under the jamming attack.
The main contributions of this work can be summarized as follows:
\begin{itemize}
    \item  In a cooperative scenario, we propose two estimation methods for finding the direction of the jamming channel vector without needing the complete CSI of JNs.  Unlike several works in the literature, the proposed method does not require a silent interval during the time frame for  the jamming channel estimation, making it well-suited for combating agile reactive jamming attacks. We demonstrate that, with a sufficient number of pilot symbols, the performance of AJ using the suggested estimation method is comparable to when AJ is applied with perfect CSI of the JNs.
    \item According to the results obtained from the estimation phase, a cooperative anti-jamming (CAJ) technique is developed. The proposed CAJ method is based on a linear matrix transformation, where the columns of the matrix are orthogonal to the jamming channel direction. We demonstrate that, under a strong deceptive jamming signal,  with a sufficient number of SNs, the overall performance of the proposed CAJ  is near to the case where no jamming attack exists.
    \item In addition, we extend the proposed estimation method along with the proposed CAJ method to combat the jamming attack by multiple JNs while supporting multiple friendly users.
    \item Analytical formulas for the outage probability of the proposed CAJ technique integrated with the proposed estimation methods are presented under the jamming attack. The analytical findings are in excellent agreement with the corresponding simulation results. 
    \item  Through simulations, we thoroughly investigate the impact of fast-fading and time-varying (TV) channels, due to mobile JNs with high Doppler shift,  on the performance of our proposed method. Our results demonstrate that the CAJ method remains highly robust as long as the coherence time of the jamming or friendly communication channel is not extremely short.
\end{itemize}
This study is organized as follows. The system model is presented in Section \ref{sec_sysmod}. The proposed estimation and  CAJ methods are detailed in Section \ref{propSec}. The analytical discussion and extension to multiple JNs are formulated in Sections \ref{sec_analytical} and \ref{sec_multiple_caj}, respectively. The simulation results and discussions are included in Section \ref{sec_simul}. The conclusion of this work is provided in Section \ref{sec_con}.
\section{System Model }\label{sec_sysmod}
\begin{figure}[t]
    \centering
    \includegraphics[width=.8\linewidth]{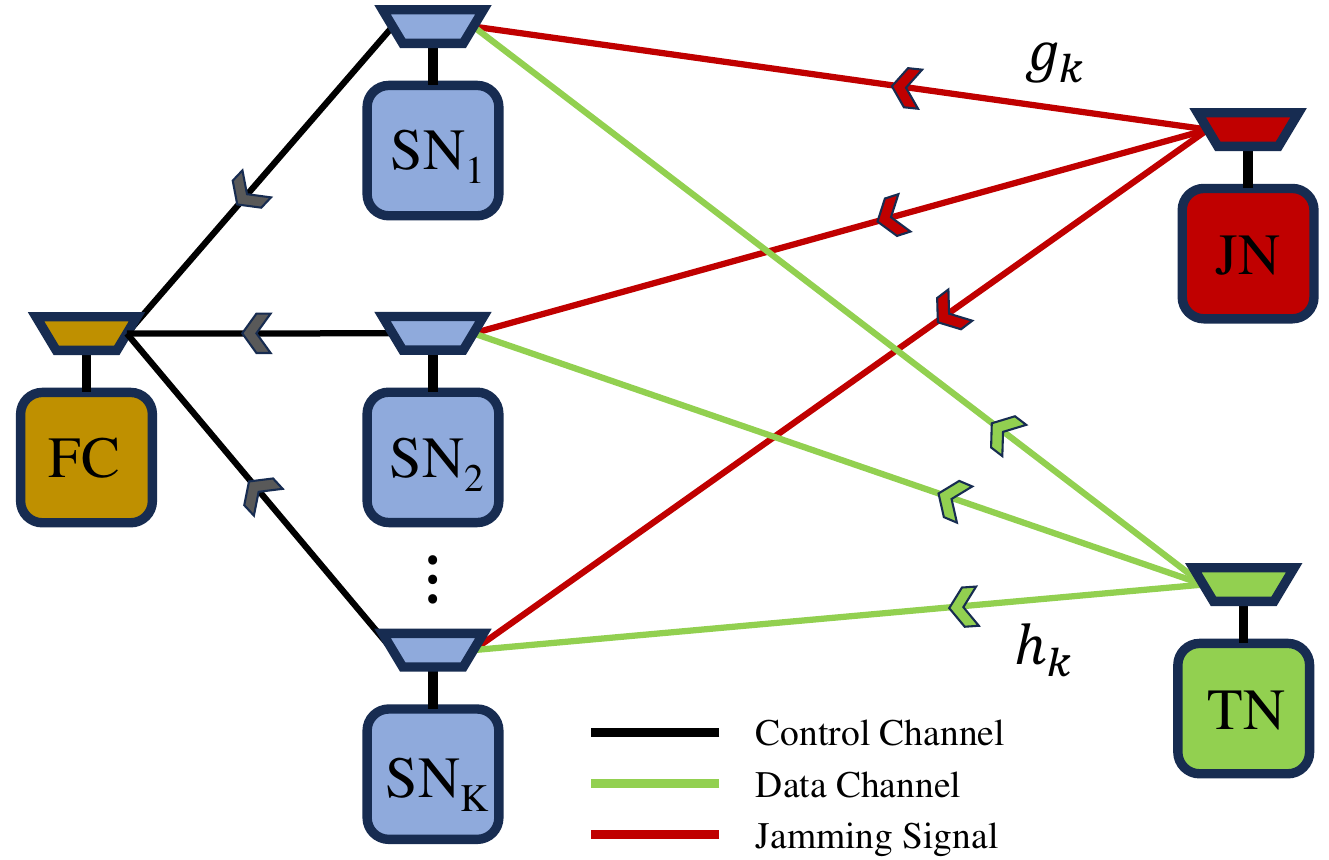}
    \caption{A network of SNs connected to the FC attempts to communicate with a friendly TN in the presence of a malicious JN.}
    \label{fig:SysMod}
\end{figure}

%
%
%
%
In this work, we consider $K$ single-antenna SNs that are subject to jamming while communicating with a single-antenna  TN. The TN transmits the $n$-th  symbol $s[n]$, forming a transmitted vector $\textbf{s}_t^T = [s[1], \dots, s[N_t]]$ over a time frame, where the superscript $T$ denotes the transpose operation. Simultaneously, a single-antenna JN transmits $\textbf{j}_t^T = [j[1], \dots, j[N_t]]$, typically using the same modulation type but with a higher power, aiming to deceive the SNs and disrupt the communication between the TN and SNs. Fig. \ref{fig:SysMod} illustrates the system model. The subscript $t$ can indicate two intervals in a time frame, $t \in \{t_p, t_d\}$. This time frame is depicted in Fig. \ref{fig0} with two intervals. During the first interval, which lasts for $t_p$, the pilot sequence is transmitted by the TN. Thus, $\textbf{s}_{t_p}$ includes $N_{t_p}$ pilot symbols that are known at the fusion center (FC). In the rest of the time frame, with a duration of $t_d$, the TN transmits the data vector $\textbf{s}_{t_d}$ with $N_{t_d}$ symbols. Thus, the received signal at the $k$-th SN during time interval $t$ is
\begin{equation} \label{sysmmodeleq1}
 \textbf{y}_{t, k} = h_k \textbf{s}_t + g_k \textbf{j}_t + \textbf{w}_{t, k},
\end{equation}
where $h_k$ is the channel between the $k$-th SN and the TN, and $g_k$ is the channel between the $k$-th SN and the JN. The noise vector at the $k$-th SN during time interval $t$ is denoted by $\textbf{w}_{t,k}$.
Similar to cooperative models in \cite{Cooperative_WED, pietra1}, we assume that the SNs send all their local data to an FC through a perfect control channel. Therefore, the FC has access to all received symbols from the $K$ SNs and organizes them into a matrix $\textbf{Y}_t = [\textbf{y}_{t,1}, \dots, \textbf{y}_{t,K}]^T$, where $\textbf{Y}_t \in \mathbb{C}^{K \times N_t}$ represents the data during time interval $t$. Using (\ref{sysmmodeleq1}), we can express the received signal matrix as 
\begin{equation}\label{eq1}
    \textbf{Y}_t =  \textbf{h} \textbf{s}_t^T + \textbf{g}\textbf{j}_t^T  + \textbf{W}_t,
\end{equation}
where $\textbf{h}= [ h_1, ...,  h_k]^T$,  $\textbf{g}= [ g_1, ...,  g_k]^T$, $\textbf{W}_t=[\textbf{w}_{t, 1}, ..., \textbf{w}_{t, K} ]^T$, and $\textbf{W}_t \in \mathbb{C}^{K \times N_t}$ . 
   The vector $\textbf{h}$ includes the channel gains between SNs and the TN, and $\textbf{g}$ denotes the channel gains vector between SNs and the JN.  Additionally, we assume that the channel vectors and sequence of signals are not identical, i.e., $\textbf{h} \neq \textbf{g}$, and $\textbf{s} \neq \textbf{j}$, which implies that $\mathrm{rank}(\textbf{h} \textbf{s}_t^T + \textbf{g}\textbf{j}_t^T )=2$.
Moreover, $ \|\textbf{s}_t\|_2= \sqrt{P_{t,s}},  \|\textbf{j}_t\|_2  = \sqrt{P_{t,j}}$,  where $\|\cdot\|_2$ is the Euclidean norm. For each column of the noise matrix, we assume that the noise is circularly symmetric complex white Gaussian (CSCWG), and we have $\textbf{w}_t[n] \sim \mathcal{CN}(0, \sigma^2 \textbf{I})$, where $\textbf{w}_t[n]$ is the $n$-th column of $\textbf{W}_t$, and shows the values of noise in all SNs at the $n$-th moment.
We also assumed that during the time frame with a duration of $t_f=t_p+t_d$ and with $N_{t_f}=N_{t_f}+N_{t_d}$ symbols, channel gains follow the slow-fading model.  The entire duration of the time frame, which lasts for $t_f$, should be much shorter than the coherence time of the channels to have fixed channel gains during the time frame. 
\section {Proposed Mehtod}\label{propSec}
\subsection{Estimation of the Direction of the Jamming Channel Vector }\label{estimation_sec}
If we obtain information about the jamming channel 
$\textbf{g}$, we can construct a matrix 
$\textbf{G}^\perp$
  whose columns span the null space of 
$\textbf{g}$, meaning they are orthogonal to 
$\textbf{g}$. By transforming the received signal matrix using 
$\textbf{G}^\perp$, we project the received signals onto this null space, effectively neutralizing the jamming attack, regardless of the type of jamming signal. 
 As long as the jamming channel $\textbf{g}$ is known and $\textbf{G}^\perp$ is constructed accordingly, the technique can neutralize the interference, regardless of the jamming vector's characteristics, e.g., amplitude, phase, frequency, or direction of arrival. This approach remains effective even when we only estimate the direction of the jamming channel vector. It is important to note that throughout this paper,  the  term  ``direction of the jamming channel vector'' specifically refers  to 
 $\bar{\textbf{g}}$, where the jamming channel vector 
$\textbf{g}$
 can be decomposed into its direction 
$\bar{\textbf{g}}$
 and its magnitude 
$\|\textbf{g}\|_2$ as
$\textbf{g}=\bar{\textbf{g}}\|\textbf{g}\|_2$. In what follows, we proposed two methods to estimate  $\bar{\textbf{g}}$.
\begin{figure}[t]
\centering

\begin{tikzpicture}
\draw [line width=2pt] (0,0)--(.5,0) --( .5,1) -- (0,1);

\fill[blue!30] (.5,0) rectangle +(1.5,1);
\node  at (1.2,0.5) {Pilots};
\fill[green!30] (2,0) rectangle +(6,1);
\node  at (4.9,0.5) {Data};
\node  at (8.4,0.5) {{\large...}};
\node  at (.1,0.5) {{\large...}};

\draw[line width=2pt] (0.5,0) rectangle +(7.5,1);
\draw[line width=2pt] (.5,0) rectangle +(1.5,1);
\draw[line width=1.5pt,<->] (0.5, -0.25) -- (2,-0.25);
\draw[line width=1.5pt,<->] (0.5, 1.2) -- (8,1.25);

\draw[line width=1.5pt,<->] (2, -0.25) -- (8,-0.25);
\node  at (3.9,1.5) {$t_f$};
\node  at (1.2,-0.5) {$t_p$};
\node  at (4.9,-0.5) {$t_d$};
\draw [line width=2pt] (8.5,0)--(8,0) -- (8,1) -- (8.5,1);

\end{tikzpicture}
\caption{The time frame with two intervals of pilot transmission and data transmission by the TN.}\label{fig0}
\end{figure}
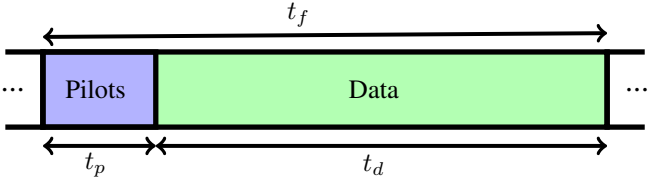  
\subsubsection{Normalized Least Squares}
During $t_p$, since $\textbf{s}_{t_p}$ is known at the SNs, we can find an orthogonal vector to the transmitted vector of pilots, denoted by $\textbf{s}^{\perp}_{t_p}$,
where $\textbf{s}^{H}_{t_p}\textbf{s}^{\perp}_{t_p}=0$, and $\| \textbf{s}^{\perp}_{t_p}\|_2=1$.
By projecting the received matrix during $t_p$ onto $\textbf{s}^{\perp}_{t_p}$, we have
\begin{equation}\label{eq2nn}
    \textbf{Y}_{t_p} {\textbf{s}^{\perp}_{t_p}}^{*} = \textbf{h} \textbf{s}_{t_p}^{T} {\textbf{s}^{\perp}_{t_p}}^{*} + \textbf{g}\textbf{j}_{t_p}^{T}{\textbf{s}^{\perp}_{t_p}}^{*}+ \textbf{W}_{t_p}{\textbf{s}^{\perp}_{t_p}}^{*},
\end{equation}
where $\textbf{Y}_{t_p} {\textbf{s}^{\perp}_{t_p}}^{*} \in \mathbb{C}^{K \times 1}$, and $*$ denotes the conjugate.
Using the orthogonality of  $\textbf{s}^{H}_{t_p}$ and $\textbf{\textbf{s}}^{\perp}_{t_p}$, we have
\begin{equation}\label{eq3nn}
     \textbf{Y}_{t_p} {\textbf{s}^{\perp}_{t_p}}^{*} =(\textbf{j}_{t_p}^{T}{\textbf{s}^{\perp}_{t_p}}^{*})\textbf{g}+ \textbf{W}_{t_p}{\textbf{s}^{\perp}_{t_p}}^{*}=
     \alpha_{t_p}\textbf{g}+\textbf{w}_{t_p},
\end{equation}
where $ \alpha_{t_p}=\textbf{j}_{t_p}^{T}{\textbf{s}^{\perp}_{t_p}}^{*}$ and $\textbf{W}_{t_p}{\textbf{s}^{\perp}_{t_p}}^{*}=\textbf{w}_{t_p} \sim \mathcal{CN}(0, \sigma^2 \textbf{I})$.
Since we are interested in the direction of $\textbf{g}$, there is no need to estimate $\alpha_{t_p}$ and the magnitude of $\textbf{g}$ separately. Therefore, using the least squares (LS) estimation \cite{Kay97}, we  can estimate the first term on the right-hand side of (\ref{eq3nn}) as 
\begin{equation}\label{eq4nn}
   \widehat{\alpha_{t_p}\textbf{g}}= \textbf{Y}_{t_p} {\textbf{s}^{\perp}_{t_p}}^{*}.
\end{equation}
It is important to note that only the magnitude of the estimated vector changes with time, and its direction is fixed during a time frame since from (\ref{eq3nn}), it is clear that the magnitude is a function of $\textbf{j}_{t_p}$ and the direction is a function of $\textbf{g}$, which is fixed during a time frame. Thus, the normalized vector is
\begin{equation}\label{eq5nn}
   \widehat{\overline{\textbf{g}}}= \overline{\textbf{Y}_{t_p} {\textbf{s}^{\perp}_{t_p}}^{*}},
\end{equation}
where the bar line denotes the normalization, i.e., $\overline{\textbf{g}}=\textbf{g}/\|\textbf{g}\|_2$.
In what follows, we refer to the estimation approach in (\ref{eq5nn}) as normalized least squares (NLS) estimation.
The received signal-to-noise ratio (SNR) from the JN in this phase can offer valuable insights into the estimation error. To proceed, we write the received SNR from the JN after removing the desired signal using (\ref{eq4nn}) as
\begin{equation}\label{eq4n}
\gamma_{r,j}=\frac{|\alpha_{t_p}|^2\|\textbf{g}\|_2^2}{K\sigma^2}=\frac{|\textbf{j}_{t_p}^H{\textbf{s}_{t_p}^{\perp}}|^2\|\textbf{g}\|_2^2}{K\sigma^2}.
\end{equation}
We can observe that $ \gamma_{r,j}$ is heavily dependent on the projection of the jamming vector onto $\textbf{s}_{t_p}^\perp$. If only a small fraction of the energy of $\textbf{j}_{t_p}$ aligns with the direction of $\textbf{s}_{t_p}^\perp$,  $\gamma_{r,j}$ will tend towards zero, resulting in a large estimation error. Conversely, when the directions of vectors $\textbf{j}_{t_p}$  and $\textbf{s}_{t_p}^\perp$ converge, the received SNR from the JN increases, leading to a more accurate estimation of the jamming channel's direction. Importantly, the NLS method functions without requiring data on $\textbf{j}_{t_p}$, which renders it practical for dealing with jamming attacks. 
\subsubsection{Estimation of Direction of JN's Channel with Eigenvectors }\label{sub_EVE}
To eliminate the signal from the TN during $t=t_p$, we can utilize multiple orthogonal vectors rather than relying on a single vector orthogonal to $\textbf{s}_{t_p}$. The space $\mathbb{C}^{N_{t_p} \times 1}$ can be spanned by $N_{t_p}$ orthogonal vectors. If we take $\overline{\textbf{s}}_{t_p}$ as one of the vectors in the set of orthogonal basis vectors for $\mathbb{C}^{N_{t_p} \times 1}$, then this space can be represented by an orthogonal basis consisting of $\{\overline{\textbf{s}}_{t_p}, \boldsymbol{\psi}_{1}, \dots,  \boldsymbol{\psi}_{N_{t_p} - 1} \}$, where each vector is mutually orthogonal and has a unit magnitude.
 By excluding $\overline{\textbf{s}}_{t_p}$, we form the matrix $\textbf{S}^\perp=[\boldsymbol{\psi}_{1}, \dots,  \boldsymbol{\psi}_{N_{t_p} - 1} ]$,  where  $\textbf{S}^\perp \in \mathbb{C}^{N_{t_p}\times (N_{t_p}-1)}$. Since  $\overline{\textbf{s}}^H_{t_p}\boldsymbol{\psi}_n=0$ for all $n=1, \dots, N_{t_p}-1$, we conclude that $\overline{\textbf{s}}^H_{t_p}\textbf{S}^\perp=\textbf{0}$. Now, using (\ref{eq1}), we can remove the desired signal by multiplying the received signal by 
$\textbf{S}^\perp$ as
\begin{equation}\label{eq7n}
\textbf{Y}_{t_p}{\textbf{S}^\perp}^*=\textbf{h} \textbf{s}_{t_p}^{T} {\textbf{S}^\perp}^* + \textbf{g}\textbf{j}_{t_p}^{T}{\textbf{S}^\perp}^*+ \textbf{W}_{t_p}{\textbf{S}^\perp}^*=\textbf{g}\textbf{a}^T_{t_p}+\textbf{N}_{t_p},
\end{equation}
where
$  \textbf{Z}_{t_p}=\textbf{Y}_{t_p}{\textbf{S}^\perp}^* $, $\textbf{N}_{t_p}=\textbf{W}_{t_p}{\textbf{S}^\perp}^*$, $\textbf{a}_{t_p}={\textbf{S}^\perp}^H\textbf{j}_{t_p}$, and $\textbf{Z}_{t_p}, \textbf{N}_{t_p}\in \mathbb{C}^{K\times (N_{t_p}-1)}$. We also used  $\textbf{s}_{t_p}^H\textbf{S}^\perp=\textbf{0}$ to derive the expression on the right-hand side of  (\ref{eq7n}). In addition, since the columns of $\textbf{S}^\perp$ are normalized, columns of $\textbf{N}_{t_p}$ follow $\mathcal{CN}(0,\sigma^2\textbf{I})$. In this case, the received SNR from the JN is
\begin{equation}\label{eq8n}
\gamma_{r,j}=\frac{\|\textbf{g}\textbf{a}^T_{t_p}\|_F^2}{K(N_{t_p}-1)\sigma^2}=\frac{\sum_{n=1}^{N_{t_p}-1}|\textbf{j}_{t_p}^H{ \boldsymbol{\psi}_{n}}|^2\|\textbf{g}\|_2^2}{K(N_{t_p}-1)\sigma^2},
\end{equation}
where $\|\cdot\|_F$ is the Frobenius norm of a matrix. In equation (\ref{eq8n}), we used the fact that  $\|\textbf{g}\textbf{a}^T_{t_p}\|_F^2=\mathrm{tr}(\textbf{a}^*_{t_p}\textbf{g}^H\textbf{g}\textbf{a}^T_{t_p})=\|\textbf{g}\|_2^2\|\textbf{a}^H_{t_p}\|^2_2$, and $\|\textbf{a}^H_{t_p}\|^2_2=\sum_{n=1}^{N_{t_p}-1}|\textbf{j}_{t_p}^H{ \boldsymbol{\psi}_{n}}|^2$, where $\mathrm{tr}(\cdot)$ is the trace of a matrix. Based on the spanning orthogonal basis, we have $P_{t,j}=\|\textbf{j}_{t_p}\|_2^2=|\textbf{j}_{t_p}^H\overline{\textbf{s}}_{t_p}|^2+\sum_{n=1}^{N_{t_p}-1}|\textbf{j}_{t_p}^H{ \boldsymbol{\psi}_{n}}|^2$. Now, we further simplify  (\ref{eq8n}) as
\begin{equation}\label{eq9n}
\gamma_{r,j}=\frac{(P_{t,j}-|\textbf{j}_{t_p}^H\overline{\textbf{s}}_{t_p}|^2)\|\textbf{g}\|_2^2}{K(N_{t_p}-1)\sigma^2}.
\end{equation}
It is important to note that in contrast to the received SNR of the NLS method in (\ref{eq4n}), the received SNR in  (\ref{eq9n}) includes all the power of the signal from the JN except for the portion of the jamming power which is in the direction of $\overline{\textbf{s}}_{t_p}$. \par
Now, by assuming $K<N_{t_p}$, we can write the singular value decomposition (SVD) for $\textbf{Z}_{t_p}$ as
\begin{equation}\label{eq10new}
    \textbf{Z}_{t_p}=\sum_{i=1}^K \lambda_i\textbf{u}_i \textbf{v}_i^H,
\end{equation}
where $\textbf{u}_i$ and $\textbf{v}_i$ are the eigenvectors, and  $\lambda_i$ are the ordered singular values with $\lambda_1\ge ... \ge \lambda_K$.
If we assume that the transmitted SNR by the JN, denoted as $\gamma_{t,j}$, is large, this implies that  $\textbf{Z}_{t_p}\approx \textbf{g}\textbf{a}_{t_p}^T$ as indicated in equation (\ref{eq7n}). Consequently,  $\textbf{Z}_{t_p}$ can be approximated as a rank-one matrix, a technique that has been utilized in various existing studies \cite{Taherpour,rank_one_haung,Mehrabian}.
This means that it has only one large singular value, and the other singular values tend to zero $\lambda_i\rightarrow 0$, where $i=2,..., K$. Thus, using  SVD in (\ref{eq10new}), we have $\textbf{Z}_{t_p}\approx \lambda_1\textbf{u}_1 \textbf{v}_1^H=  \overline{\textbf{g}}\|\textbf{g}\|_2\textbf{a}_{t_p}^T$. As a result, we can write $ \textbf{u}_1 \textbf{v}_1^H -  \overline{\textbf{g}}\textbf{a}_{t_p}^T\|\textbf{g}\|_2/\lambda_1=\textbf{0}$, where every column should be equal to zero vector, i.e., $ \textbf{u}_1 [\textbf{v}_1^*]_n -   \overline{\textbf{g}}[\textbf{a}_{t_p}]_n\|\textbf{g}\|_2/\lambda_1=\textbf{0}$, for $n=1, ..., N_{t_p}-1$. Here, $[\textbf{x}]_n$ denotes the $n$-th element of vector $\textbf{x}$. Accordingly, since $\textbf{v}_1,\textbf{a}_{t_p}\ne \textbf{0}$,  we can estimate the direction of $\textbf{g}$  as
\begin{equation}\label{eq10n}
\widehat{\overline{\textbf{g}}}=\textbf{u}_1,
\end{equation}
which we refer to as eigenvector (EV) estimation. Having an estimation of $\overline{\textbf{g}}$, we can proceed to the next step to reduce the impact of jamming on the desired signal. \par
Interestingly, to find $\widehat{\overline{\textbf{g}}}$,  using (\ref{eq7n}) and (\ref{eq10new}), we can write $\widehat{\bar{\textbf{g}}}\approx\textbf{Z}_{t_p}\textbf{v}_1/\lambda_1=\textbf{g}\textbf{a}^T_{t_p}\textbf{v}_1/\lambda_1+\textbf{N}_{t_p}\textbf{v}_1/\lambda_1$. Thus,  the received SNR from the JN becomes
\begin{equation}\label{newSNR}
    \gamma_{r,j} = \frac{\|\textbf{g}\textbf{a}^T_{t_p}\textbf{v}_1\|^2_2}{K\sigma^2}\approx\frac{\|\textbf{g}\|^2_2\|\textbf{a}_{t_p}\|^2_2}{K\sigma^2}=\frac{(P_{t,j}-|\textbf{j}_{t_p}^H\overline{\textbf{s}}_{t_p}|^2)\|\textbf{g}\|_2^2}{K\sigma^2},
\end{equation}
where using the rank-one property and $\|\textbf{g}\|_2\|\textbf{a}_{t_p}\|_2\bar{\textbf{g}}\bar{\textbf{a}}_{t_p}^T\approx \lambda_1\textbf{u}_1\textbf{v}_1^H$, we assumed $\bar{\textbf{a}}_{t_p}\approx \textbf{v}_1^*$. By comparing $\gamma_{r,j}$ for the NLS and EV methods in equations (\ref{eq4n}) and (\ref{newSNR}), respectively, we observe that $\gamma_{t,j}$ can be significantly larger for the EV method. This enhancement is attributed to its ability to capture a substantially larger portion of the jamming power $P_{t,j}$, resulting in an improved estimation performance.
\subsection{Counteracting the Jamming Signal}\label{sec.CAJ}
After estimating the direction of the jamming channel vector, we can remove a great portion of the jamming signal. The jamming channel vector spans a line (one-dimensional space) in $\mathbb{C}^{K\times1}$. The simplest method is to find an orthogonal vector to $\overline{\textbf{g}}$ and project the received matrix onto this vector to eliminate the jamming signal. However, in order to preserve the DoFs stemming from the multiple SNs, we use the orthogonal complement space of the line spanned by $\overline{\textbf{g}}$. If we consider the orthogonal basis for $\mathbb{C}^{K\times1}$ as $\{\widehat{\overline{\textbf{g}}}, \textbf{g}_{1} ^{\perp},..., \textbf{g}_{K-1} ^{\perp}\}$, we can form a matrix  $\textbf{G}^\perp \in \mathbb{C}^{K\times (K-1)}$, given by $\textbf{G}^\perp=[\textbf{g}_{1}^{\perp}, ..., \textbf{g}_{K-1}^{\perp}]$. Now, introducing  $\textbf{E}_{t_p}={\textbf{G}^\perp}^H\textbf{W}_{t_p}+\textbf{E}_{g}$ and $\textbf{E}_g={\textbf{G}^\perp}^H\textbf{g}\textbf{j}_{t_p}^T$, the received matrix in (\ref{eq1}) is transformed  by the matrix $\textbf{G}^\perp$ as 
\begin{equation}\label{eq11n}
{\textbf{G}^\perp}^H\textbf{Y}_{t_p}= {\textbf{G}^\perp}^H \textbf{h} \textbf{s}_{t_p}^T + \textbf{E}_{t_p}.
\end{equation}
 The matrix $\textbf{E}_g$ arises due to errors in the estimation of the jamming channel. We also assume that the columns of  $\textbf{E}_{t_p}$, denoted by $\textbf{e}_{t_p}[n]$, follow a complex normal distribution as $\textbf{e}_{t_p}[n]\sim \mathcal{CN}(\textbf{0},\sigma^2_e\textbf{I})$.\par
After the transformation with $\textbf{G}^\perp$, we need to estimate the new transformed channel.
By introducing $\textbf{f}={\textbf{G}^\perp}^H \textbf{h}$, we have
\begin{equation}\label{eq2n_rev}
{\textbf{G}^\perp}^H\textbf{Y}_{t_p}= \textbf{f} \textbf{s}_{t_p}^T + \textbf{E}_{t_p},
\end{equation}
where we use the least squares (LS) estimation for channel estimation \cite{Kay97}. As the pilot vector is known at the FC, the pseudoinverse of the pilot vector $\textbf{s}_{t_p}^T$ is $\textbf{s}_{t_p}^+=\textbf{s}_{t_p}^*||\textbf{s}_{t_p}||^{-2}_2$.  We multiply (\ref{eq2n_rev}) by $\textbf{s}_{t_p}^+$, and using $\textbf{s}_{t_p}^T\textbf{s}_{t_p}^+=1$, we have
\begin{equation}\label{eq3n}
{\textbf{G}^\perp}^H\textbf{Y}_{t_p}\textbf{s}_{t_p}^+=  \textbf{f} + \textbf{E}_{t_p}\textbf{s}_{t_p}^+ ,
\end{equation}
As a result, we can estimate the transformed channel   $\textbf{f}$ as
\begin{equation}\label{eq12n}
\widehat{\textbf{f}}={\textbf{G}^\perp}^H\textbf{Y}_{t_p}\textbf{s}_{t_p}^*\|\textbf{s}_{t_p}\|_2^{-2}= \textbf{f} + \textbf{E}_{t_p}\textbf{s}_{t_p}^*\|\textbf{s}_{t_p}\|_2^{-2}.
\end{equation}
\par
By estimating $\textbf{f}$ and ${\textbf{G}^\perp}$ during $t_p$, we are now able to use a detection method, such as zero-forcing (ZF) to estimate the data transmitted by the TN in $\textbf{s}_{t_d}$ during $t_d$. Thus, with ZF,  we can detect $\textbf{s}_{t_d}$  as 
\begin{equation}\label{eq13n}
\widehat{\textbf{s}}_{t_d}^T=\|\widehat{\textbf{f}}\|^{-2}_2\widehat{\textbf{f}}^H{\textbf{G}^\perp}^H\textbf{Y}_{t_d}=\|\widehat{\textbf{f}}\|^{-2}_2\widehat{\textbf{f}}^H \textbf{f}\textbf{s}_{t_d}^T + \|\widehat{\textbf{f}}\|^{-2}_2\widehat{\textbf{f}}^H\textbf{E}_{t_d}.
\end{equation}
The key idea behind the proposed AJ method is 
to employ the orthogonal space of the jamming channel $\textbf{g}$ and project the received matrix onto this space. This proposed CAJ method enables us to benefit from the diversity due to the collaboration of SNs while eliminating a large portion of the jamming signal in the orthogonal space of $\textbf{g}$. The steps required for channel estimation and jamming removal, namely the EV-CAJ and the NLS-CAJ algorithms, are briefly explained in Algorithm \ref{alg:one}.
\begin{algorithm}[t]
\caption{The NLS-CAJ and EV-CAJ algorithms}\label{alg:one}
\begin{algorithmic}[1] 
\State \textbf{Input:}\ $N_{t_p}, N_{t_d}, {t_p}, t_d$.
\State \textbf{Output:}\ $\widehat{\textbf{s}}_{t_d}$.

\State \textbf{During Pilot Block Transmission in $t_p$:}
    \State Receive the observation matrix $\textbf{Y}_{t_p}$.
    \State Form $\textbf{s}_{t_p}^\perp$ or $\textbf{S}^\perp$ using the pilot vector $\textbf{s}_{t_p}$ based on the NLS or the EV method.
    \State Remove the signal transmitted by the TN.
    \State Estimate $\overline{\textbf{g}}$ according to the EV in (\ref{eq10n}), or the NLS estimation method in (\ref{eq5nn}).
    \State Form $\textbf{G}^\perp$ using $\widehat{\overline{\textbf{g}}}$.
    \State Eliminate the signal transmitted by the JN using ${\textbf{G}^{\perp}}^H\textbf{Y}_{t_p}$.
    \State Find the estimation of the transformed channel $\widehat{\textbf{f}}$ using the pilot vector $\textbf{s}_{t_p}$ and formula (\ref{eq12n}).
  
\State \textbf{During Data Block Transmission in $t_d$:}
    \State Receive the data matrix $\textbf{Y}_{t_d}$.
    \State Eliminate the signal from the JN by ${\textbf{G}^{\perp}}^H\textbf{Y}_{t_d}$.
    \State Detect the desired signal vector $\textbf{s}_{t_d}$ employing $\widehat{\textbf{f}}$ and formula (\ref{eq13n}).

\end{algorithmic}
\end{algorithm}
%
\section{Analytical Discussion on the Proposed Methods}\label{sec_analytical}
This section analyzes the received SNR from the TN after jamming suppression to evaluate both estimation accuracy and suppression effectiveness. In contrast, Section III used the JN's SNR to assess jamming direction estimation. 
The first step is to examine the estimation error of the EV. In the estimation phase, we assume that the JN's channel direction is a deterministic parameter. The Cramer-Rao lower bound (CRLB) is a useful metric to show the accuracy of the estimation method.  To obtain the CRLB, we start with the log-likelihood function (LLF) of the observations in equation (\ref{eq7n}). Since columns of  $\textbf{N}_{t_p}$ in (\ref{eq7n}) follow a zero-mean CSCWG distribution, i.e.,  $\textbf{n}_{t_p}[n]\sim\mathcal{CN}(0,\sigma^2\textbf{I})$,  the columns of the observation matrix also follow a CSCWG distribution as $\textbf{z}_{t_p}[n]\sim\mathcal{CN}(\textbf{g}[\textbf{a}_{t_p}]_n,\sigma^2\textbf{I})$. We also employ the notation $[n]$ in $\textbf{z}_{t_p}{[n]}$ and $\textbf{n}_{t_p}{[n]}$, which are the $n$-th columns of $\textbf{Z}_{t_p}$ and $\textbf{N}_{t_p}$, to represent their dependence on time.
 In addition,
$[\textbf{a}_{t_p}]_n$ shows the $n$-th element of the vector $\textbf{a}_{t_p}$. The LLF of observations $\textbf{Z}_{t_p}$ is
\begin{equation}\label{eq14n}
L(\textbf{Z}_{t_p})=c_o-\sigma^{-2}\sum_{n=1}^{N_{t_p}-1}\| \textbf{z}_{t_p}[n]-\textbf{g}[\textbf{a}_{t_p}]_n\|^2_2,
\end{equation}
where  $c_0$ is a constant that is independent of $\textbf{g}$. To find the CRLB, we can use the following  formula to compute the Fisher information matrix (FIM) \cite{Kay_Rao_Complex}\cite{Kay_CRLB_Complex}
\begin{equation}\label{eq15n}
\textbf{I}(\underline{\textbf{g}})
=\mathrm{E} \Big\{ \frac{\partial L(\textbf{Z}_{t_p})}{\partial \underline{\textbf{g}}^*} \frac{\partial L(\textbf{Z}_{t_p})^H}{\partial \underline{\textbf{g}}^*}  \Big\},
\end{equation}
where  $\mathrm{E}\{\cdot\}$ is the expectation, and $\underline{\textbf{g}}=[\textbf{g}^T \textbf{g}^H]^T$. Therefore, we can rewrite the FIM as 
\begin{equation}\label{eq16n}
\begin{split}
\textbf{I}(\underline{\textbf{g}})
&=\mathrm{E} \Bigg\{ \Bigg[\begin{array}{c}
     \frac{{\displaystyle \partial L(\textbf{Z}_{t_p})}}{{\displaystyle \partial \textbf{g}^*}} \\
     \frac{{\displaystyle\partial L(\textbf{Z}_{t_p})}}{{\displaystyle \partial \textbf{g}}}
\end{array} \Bigg] \Big[\frac{\partial L(\textbf{Z}_{t_p})^H}{\partial \textbf{g}^*}  \frac{\partial L(\textbf{Z}_{t_p})^H}{\partial \textbf{g}}\Big] \Bigg\}\\
&=\Bigg[\begin{array}{cc}
     \Tilde{\textbf{I}}(\textbf{g}) & \Tilde{\textbf{J}}(\textbf{g})\\
     \Tilde{\textbf{J}}(\textbf{g})^* & \Tilde{\textbf{I}}(\textbf{g})^*
\end{array} \Bigg].
\end{split}
\end{equation}
The FIM matrix can be determined by finding the four block matrices in (\ref{eq16n}). Employing properties of the derivative of a scalar by a vector, we have \cite{Kay_CRLB_Complex}
\begin{equation}\label{eq17n}
\begin{split}
\frac{\partial L(\textbf{Z}_{t_p})}{\partial \textbf{g}^*} =\sigma^{-2}\sum_{n=1}^{N_{t_p}-1}[\textbf{a}_{t_p}]_n^*( \textbf{z}_{t_p}[n]-\textbf{g}[\textbf{a}_{t_p}]_n),
 \\
\frac{\partial L(\textbf{Z}_{t_p})}{\partial \textbf{g}} =\sigma^{-2}\sum_{n=1}^{N_{t_p}-1}[\textbf{a}_{t_p}]_n( \textbf{z}_{t_p}[n]-\textbf{g}[\textbf{a}_{t_p}]_n)^*.
\end{split}
\end{equation}
$\Tilde{\textbf{I}}(\textbf{g})$ is the upper left block in (\ref{eq16n}), which can be written as 
\begin{equation}\label{eq18n}
\begin{split}
\Tilde{\textbf{I}}(\textbf{g})=\mathrm{E} \Bigg\{ \frac{\partial L(\textbf{Z}_{t_p})}{\partial \textbf{g}^*}  \frac{\partial L(\textbf{Z}_{t_p})^H}{\partial \textbf{g}^*}\Bigg\}.
\end{split}
\end{equation}
We plug (\ref{eq17n}) into (\ref{eq18n}) which results in
\begin{equation}\label{eq19n}
\begin{split}
\Tilde{\textbf{I}}(\textbf{g})&=
\frac{1}{\sigma^{4}}\sum_{n=1}^{N_{t_p}-1}\sum_{m=1}^{N_{t_p}-1}[\textbf{a}_{t_p}]_n^*[\textbf{a}_{t_p}]_m\\ &\times \mathrm{E} \big\{ ( \textbf{z}_{t_p}[n]-\textbf{g}[\textbf{a}_{t_p}]_n)( \textbf{z}_{t_p}[m]-\textbf{g}[\textbf{a}_{t_p}]_m)^H\big\}.
\end{split}
\end{equation}
Due to the statistical independence and zero-mean property of the noise, we conclude that $\mathrm{E}\big\{ \textbf{n}_{t_p}[n] \textbf{n}_{t_p}[m]^H\big\}= \delta[n-m]\sigma^2\textbf{I}$, where $\delta[x]$ is the discrete delta function, $\delta[x]=0$ for $x\ne 0$, and $\delta[x]=1$ for $x=0$. According to this property of the noise vector,  we are able to further simplify the equation in (\ref{eq19n}) as
\begin{equation}\label{eq20n}
\begin{split}
\Tilde{\textbf{I}}(\textbf{g})=
\frac{1}{\sigma^{2}}\sum_{n=1}^{N_{t_p}-1}|[\textbf{a}_{t_p}]_n|^2\textbf{I}=\frac{1}{\sigma^{2}}\sum_{n=1}^{N_{t_p}-1}|\textbf{j}_{t_p}^H{ \boldsymbol{\psi}_{n}}|^2\textbf{I}.
\end{split}
\end{equation}
Based on (\ref{eq16n}), to compute $\Tilde{\textbf{J}}(\textbf{g})$, we should find the following expectation
\begin{equation}\label{eq21n}
\begin{split}
\Tilde{\textbf{J}}(\textbf{g})=\mathrm{E} \Bigg\{ \frac{\partial L(\textbf{Z}_{t_p})}{\partial \textbf{g}^*}  \frac{\partial L(\textbf{Z}_{t_p})^H}{\partial \textbf{g}}\Bigg\},
\end{split}
\end{equation}
and by plugging (\ref{eq17n}) into (\ref{eq21n}), we get
\begin{equation}\label{eq22n}
\begin{split}
\Tilde{\textbf{J}}(\textbf{g})&=
\frac{1}{\sigma^{4}}\sum_{n=1}^{N_{t_p}-1}\sum_{m=1}^{N_{t_p}-1}[\textbf{a}_{t_p}]_n^*[\textbf{a}_{t_p}]_m^*\\ &\times \mathrm{E} \big\{ ( \textbf{z}_{t_p}[n]-\textbf{g}[\textbf{a}_{t_p}]_n)( \textbf{z}_{t_p}[m]-\textbf{g}[\textbf{a}_{t_p}]_m)^T\big\},
\end{split}
\end{equation}
where similar to the argument used for the simplification of (\ref{eq19n}), we can conclude that the expected term in (\ref{eq22n}) is equal to zero for $n\ne m$. Additionally, for  $n=m$, we have $\mathrm{E}\big\{ \textbf{n}_{t_p}[n] \textbf{n}_{t_p}[n]^T\big\}=\textbf{0}$, since we assume that the noise samples are circularly symmetric. This implies that $\Tilde{\textbf{J}}(\textbf{g})=\textbf{0}$, and $\textbf{I}(\underline{\textbf{g}})$ is a diagonal matrix where its diagonal elements can be determined using (\ref{eq20n}).
Thus, based on the CRLB, for the  covariance of every unbiased estimator like $\hat{\textbf{g}}$, we have
\begin{equation}\label{eq23n}
[\textbf{C}_{\hat{\textbf{g}}}]_{ii}\ge [\Tilde{\textbf{I}}(\textbf{g})^{-1}]_{ii},
\end{equation}
where for the efficient estimator, which has the best performance, the equality holds \cite{Kay_CRLB_Complex}. In (\ref{eq23n}), $[\textbf{X}]_{ij}$ shows the element in the $i$-th row and $j$-th column of matrix $\textbf{X}$.
If we approximate the performance of the EV with the performance of the efficient estimator, we can write the covariance of the EV as $\textbf{C}_{\widehat{\overline{\textbf{g}}}}=\Tilde{\textbf{I}}(\textbf{g})^{-1}/K=\sigma^2_{\epsilon}\textbf{I}$, where $\sigma^2_{\epsilon}=\sigma^2/K\sum_{n=1}^{N_{t_p}-1}|\textbf{j}_{t_p}^H{ \boldsymbol{\psi}_{n}}|^2$, and for simplicity, we assumed that $\|\textbf{g}\|_2=\sqrt{K}$ and $\overline{\textbf{g}}=\textbf{g}/\sqrt{K}$. In addition, since the EV method is an unbiased estimator that is a linear function of the observation matrix  $\textbf{Z}_{t_p}$, we have $\widehat{\overline{\textbf{g}}} \sim \mathcal{CN}(\overline{\textbf{g}},\sigma^2_{\epsilon}\textbf{I})$ \cite{kay2009fundamentals}. Consequently, the estimation error for the EV method is $\textbf{e}=\widehat{\overline{\textbf{g}}}-\overline{\textbf{g}}$, where $\textbf{e}\sim \mathcal{CN}(\textbf{0},\sigma^2_{\epsilon}\textbf{I})$. We can now integrate the error in the estimation of the JN's channel direction during the jamming removal phase. In the best-case scenario, the desirable $\textbf{G}^\perp$ satisfies ${\textbf{G}^\perp}^H\overline{\textbf{g}}=\textbf{0}$. However, since we use the estimation of the JN's channel direction, we have to consider the estimation error. Therefore, we can write ${\textbf{G}^\perp}^H\widehat{\overline{\textbf{g}}}={\textbf{G}^\perp}^H\overline{\textbf{g}}+{\textbf{G}^\perp}^H\textbf{e}=\textbf{0}$, and consequently, it can be concluded that 
\begin{equation}\label{eq24n}
 {\textbf{G}^\perp}^H\overline{\textbf{g}}=-{\textbf{G}^\perp}^H\textbf{e}. 
\end{equation}
Evidently,  as columns of $\textbf{G}^\perp$ are normalized, the transformed vector of estimation error also follows $-{\textbf{G}^\perp}^H\textbf{e}\sim \mathcal{CN}(\textbf{0},\sigma^2_{\epsilon}\textbf{I})$, while its size is $K-1$.
Using (\ref{eq1}) and its transformed version with ${\textbf{G}^\perp}$, we have
\begin{equation}\label{eq25n}
    {\textbf{G}^\perp}^H\textbf{Y}_t =  \textbf{f} \textbf{s}_t^T -\sqrt{K}{\textbf{G}^\perp}^H\textbf{e}\textbf{j}_t^T + {\textbf{G}^\perp}^H\textbf{W}_t,
\end{equation}
which is obtained using (\ref{eq24n}), $\|\textbf{g}\|=\sqrt{K}$, and $\textbf{f}={\textbf{G}^\perp}^H\textbf{h}$. 
Based on (\ref{eq25n}),
we can formulate the received SNR from the TN after  the jamming removal as
\begin{equation}\label{eq26n}
\gamma_{r,s}=\frac{\|\textbf{f}\|_2^2P_{t,s}}{(K-1)(K\sigma^2_{\epsilon}P_{t,j}+\sigma_2)},
\end{equation}
where $P_{t,s}=\|\textbf{s}_t\|_2^2$, and $P_{t,j}=\|\textbf{j}_t\|_2^2$. Plugging $\sigma^2_{\epsilon}=\sigma^2/K\sum_{n=1}^{N_{t_p}-1}|\textbf{j}_{t_p}^H{ \boldsymbol{\psi}_{n}}|^2$ into (\ref{eq26n}), we get
\begin{equation}\label{eq27n}
\gamma_{r,s}=\frac{\|\textbf{f}\|_2^2P_{t,s}}{\sigma_2(K-1)(\frac{P_{t,j}}{\sum_{n=1}^{N_{t_p}-1}|\textbf{j}_{t_p}^H{ \boldsymbol{\psi}_{n}}|^2}+1)}=\|\textbf{f}\|_2^2\beta.
\end{equation}\par
This analytical formula for the received SNR after the removal of the JN's signal,  enables us to calculate the outage probability as $P_{out}=\mathrm{Pr}\{\gamma_{r,s}<\gamma_{th}\}$. By assuming that the channel vector between the TN and SNs follows a  Rayleigh fading model and a normal distribution with $\textbf{h}\sim\mathcal{CN}(\textbf{0},\textbf{I})$, we can 
 argue that $\|\textbf{f}\|_2^2$ follows a chi-squared distribution with $2(K-1)$ real DoF. Denoting the cumulative distribution function  (CDF) of  a real-valued random variable with chi-squared distribution and  with $k$ real DoF by $\chi^2_k(.)$, the outage probability for the  EV-CAJ is 
\begin{equation}\label{eq28n}
    P_{out}=\mathrm{Pr}\{\beta\|\textbf{f}\|^2_2<\gamma_{th}\}=\chi^2_{2(K-1)}(2\gamma_{th}/\beta).
\end{equation}\par
This analytical formula for the outage probability of the EV-CAJ scheme can be applied to the NLS-CAJ  method to find its outage probability since the CAJ method is common in both approaches. The only difference is that one orthogonal vector is used in the NLS estimation for the removal of the TN's signal. Thus, $\beta$ in (\ref{eq27n}) for the NLS method becomes
\begin{equation}
\beta=\frac{P_{t,s}}{\sigma_2(K-1)(\frac{P_{t,j}}{|\textbf{j}_{t_p}^H{ \textbf{s}_{t_p}^\perp}|^2}+1)}.
\end{equation}\par
In the following, we will show the excellent agreement of this analytical expression for outage probability with  Monte Carlo (MC) simulations. 

\section{CAJ for multiple JNs and Multiple TNs}\label{sec_multiple_caj}
In this section, we extend the proposed CAJ method to the case where multiple JNs attempt to disrupt the communication of the network with $K_t$ friendly TNs. Similar to (\ref{eq1}), the received symbols from $K$ SNs at the FC  during time interval $t \in \{t_p,t_d\}$  in the presence of $K_j$ JNs are given by  
 \begin{equation}\label{eq29new}
    \textbf{Y}_{t} = \sum_{i=1}^{K_t} \textbf{h}_i \textbf{s}_{t,i}^T + \sum_{l=1}^{K_j}\textbf{g}_l\textbf{j}_{t,l}^T  + \textbf{W}_{t},
\end{equation}
 where
the vectors $\textbf{h}_i$ and $\textbf{g}_l$  include channel gains between SNs and the $i$-th TN and the $l$-th JN, respectively.  $\textbf{s}_{t,i}$ and $\textbf{j}_{t,l}$ are symbol vectors transmitted by the $i$-th TN and the $l$-th JN, respectively. We assume that JNs employ the same modulation type used by the TNs to deceive SNs. \par
During  pilot transmission at $t = t_p$ with $N_{t_p}$ symbols, each TN uses a different pilot vector $\textbf{s}_{t_p, i}$, $i=1, \dots, K_t$, where these vectors are pairwise orthogonal. Since we assume $N_{t_p} > K_t$, there still exist $N_{t_p} - K_t$ vectors, denoted by $\boldsymbol{\psi}_{n}$, that are pairwise orthogonal and are orthogonal to the pilot vectors. These vectors, along with the pilot vectors, span $\mathbb{C}^{N_{t_p} \times 1}$.
Therefore, similar to the method presented in Section \ref{sub_EVE}, a matrix whose columns are all orthogonal to the $K_t$ pilot vectors used by the TNs can be denoted as $\textbf{S}^\perp = [ \boldsymbol{\psi}_{1}, \dots, \boldsymbol{\psi}_{N_{t_p} - K_t} ]$.
 The columns of $\textbf{S}^\perp$ are normalized and are orthogonal to the subspace spanned by the set of all pilot vectors $\{\textbf{s}_{t_p,1}, \dots, \textbf{s}_{t_p,K_t}\}$. In the following, we also employ the orthogonality of pilots to each other $\textbf{s}_{t_p,i}^H\textbf{s}_{t_p,j}=0$ during channel estimation. Thus, during $t_p$, we use $\textbf{S}^\perp$ to eliminate the friendly signals  transmitted by TNs
\begin{equation}\label{eq30nn}
    \textbf{Y}_{t_p} {\textbf{S}^{\perp}}^{*} = \sum_{i=1}^{K_t} \textbf{h}_i \textbf{s}_{t_p,i}^T
    {\textbf{S}^{\perp}}^{*} + \sum_{l=1}^{K_j}\textbf{g}_l\textbf{j}_{t_p,l}^T 
    {\textbf{S}^{\perp}}^{*}+\textbf{W}_{t_p}{\textbf{S}^{\perp}}^{*}.
\end{equation}
Given that $\textbf{s}_{t_p,i}^{T} {\textbf{S}^{\perp}}^{*}=\textbf{0}$ for $i=1, \dots, K_t$ and introducing $\textbf{a}_{t_p,l}^{T}=\textbf{j}_{t_p,l}^{T}{\textbf{S}^{\perp}}^{*}$, we can restate (\ref{eq30nn})  as
\begin{equation}\label{eq31nn}
   \textbf{Z}_{t_p}= \textbf{Y}_{t_p} {\textbf{S}^{\perp}}^{*} =  \sum_{l=1}^{K_j}\textbf{g}_l \textbf{a}_{t_p,l}^{T}+ \textbf{W}_{t_p}{\textbf{S}^{\perp}}^{*}=\textbf{A}_{t_p}+\textbf{N}_{t_p},
\end{equation}
where $\textbf{A}_{t_p}=  \sum_{l=1}^{K_j}\textbf{g}_l \textbf{a}_{t_p,l}^{T}$ and $\textbf{N}_{t_p}=\textbf{W}_{t_p}{\textbf{S}^{\perp}}^{*}$. It is apparent that $\mathrm{rank}(\textbf{A}_{t_p})=K_j$ if $K_j<K$. Thus, 
the proposed CAJ method is feasible if the number of SNs exceeds the number of JNs, and we assumed that $K_j$ is known at the FC. In this case, the maximum likelihood estimation (MLE)  for $\textbf{A}_{t_p}$, constrained on $\mathrm{rank}(\textbf{A}_{t_p})=K_j$, is 
\begin{equation}\label{eq32nn}
\widehat{\textbf{A}_{t_p}}=\sum_{i=1}^{K_j}\lambda_i\textbf{u}_i\textbf{v}_i,
\end{equation}
 where the  $i$-th singular value  of $\textbf{Z}_{t_p}$ is denoted by  $\lambda_i$ and  $\lambda_1\ge ... \ge \lambda_{N_{t_p}}$. $\textbf{u}_i$ and $\textbf{v}_i$ are the left and right eigenvectors of $\textbf{Z}_{t_p}$, respectively. The derivation of (\ref{eq32nn}) is detailed in Appendix \ref{appexA}.
Clearly,  we are more interested in a matrix $\textbf{G}^\perp$ that results in $\textbf{G}^\perp\widehat{\textbf{A}_{t_p}}=\textbf{0}$
in order to remove all JNs' signals. Therefore, one straightforward option is to form $\textbf{G}^\perp$ using the left eigenvectors of $\textbf{Z}_{t_p}$ as $\textbf{G}^\perp=[\textbf{u}_{K_j+1}, ..., \textbf{u}_K]$, where $\textbf{G}^\perp \in \mathbb{C}^{K\times (K-K_j)}$. Since $\textbf{u}_i^H\textbf{u}_j=0$ for $i\ne j$, the transformation of the 
 received matrix as ${\textbf{G}^\perp}^H\textbf{Z}_{t_p}$ will efficiently eliminate a large portion of the jamming signals. The remaining DoF or available DoF after jamming removal is $K-K_j$, which suggests that a larger number of JNs would cause an increased degradation in the performance of the proposed CAJ. Subsequently, the steps explained in Section \ref{sec.CAJ} and Algorithm \ref{alg:one} can be performed for channel estimation and demodulation of the transmitted symbols by the TNs.
 Therefore, we use $\textbf{G}^{\perp}$, to counteract the jamming signal, 
and since we assume $ {\textbf{G}^{\perp}}^H \textbf{g}\approx \textbf{0}$, we can use (\ref{eq29new}) to write
\begin{equation}\label{eq6new}
   {\textbf{G}^{\perp}}^H \textbf{Y}_{t_p} \approx   \sum_{i=1}^{K_t} \textbf{f}_i \textbf{s}_{t_p,i}^T +   {\textbf{G}^{\perp}}^H \textbf{W}_{t_p},
\end{equation}
where $\textbf{f}_i={\textbf{G}^{\perp}}^H\textbf{h}_i$.
Since the pilot vectors are known at the FC,
$\textbf{f}_i$ can be estimated
using the orthogonality of the pilot vectors such as in (\ref{eq12n}) 
\begin{equation}
   \hat{\textbf{f}}_i = {\textbf{G}^{\perp}}^*\textbf{Y}_{t_p}\textbf{s}_{t_p,i}\| \textbf{s}_{t_p,i}^*\|_2^{-2}.
\end{equation}
After jamming removal during  data transmission by employing the matrix $\textbf{G}^\perp$, a similar equation to (\ref{eq6new}) can be written during the data transmission interval  $t=t_d$ as 
\begin{equation}\label{eq8new}
   {\textbf{G}^{\perp}}^H \textbf{Y}_{t_d} \approx   \sum_{i=1}^{K_t} \textbf{f}_i \textbf{s}_{t_d,i}^T +   {\textbf{G}^{\perp}}^H \textbf{W}_{t_d}.
\end{equation}
By introducing  $\textbf{F}= [\textbf{f}_1, ..., \textbf{f}_{K_t}]$ and $\textbf{S}_{t_d}=[\textbf{s}_{t_d,1}, ..., \textbf{s}_{t_d, K_t}]$,
equation (\ref{eq8new}) can be rewritten as
\begin{equation}
{\textbf{G}^{\perp}}^H\textbf{Y}_{t_d}=\textbf{FS}_{t_d}^T
+{\textbf{G}^{\perp}}^H\textbf{W}_{t_d}.
\end{equation}
 Now, using a detection method, such as zero forcing, we can detect the  symbols matrix as
\begin{equation}
    \hat{\textbf{S}}_{t_d}^T= (\hat{\textbf{F}}^H\hat{\textbf{F}})^{-1}\hat{\textbf{F}}^H{\textbf{G}^{\perp}}^H\textbf{Y}_{t_d},
\end{equation}
where $\hat{\textbf{F}}= [\hat{\textbf{f}_1}, ..., \hat{\textbf{f}}_{K_t}]$. For $K_t$  TNs,  $\hat{\textbf{F}}, \textbf{F}\in \mathbb{C}^{(K-K_j)\times K_t}$. Thus, it is required to have $K-K_j>K_t$ to ensure that $\hat{\textbf{F}}^H\hat{\textbf{F}}$ is non-singular and invertible.

\section{Simulation Results}\label{sec_simul}
\begin{table}[t]
\centering
\small
\caption{Simulation Parameters \label{tab:param}}
\begin{tabular}{|c|c|c|}
\hline
\textbf{Parameter} & \textbf{Notation} & \textbf{Value} \\
\hline
Number of SNs & $K$ & \{4, 16\} \\
\hline
Number of JNs & $K_j$ & \{1, 2\} \\
\hline
Number of TNs & $K_t$ &  1\\
\hline
Number of Pilot Symbols & $N_{t_p}$ & \{20, 50\} \\
\hline
Number of Data Symbols & $N_{t_d}$ & 1000 \\
\hline
Number of Symbols in a time frame & $N_{t_f}$ & \begin{small} $N_{t_p}+N_{t_d}$ \end{small} \\
\hline
Transmit SNR by TNs & $\gamma_{t,s}$ & 10 dB \\
\hline
Transmit SNR by JNs & $\gamma_{t,j}$ & 40 dB \\
\hline
 Threshold SNR for Outage & $\gamma_{th}$ & -10 dB \\
\hline
 Symbol Duration & $t_s$ &  - \\
\hline
Total Pilot Transmission Interval & $t_p$ &  $t_sN_{t_p}$ \\
\hline
Total Data Transmission Interval & $t_d$ &  $t_sN_{t_d}$ \\
\hline
JN's Channel Vector Coherence Time & $t_{c,g}$ &  - \\
\hline
TN's Channel Vector Coherence Time & $t_{c,h}$ &  - \\
\hline
Ratio  of $t_d$ to  $t_{c,g}$ &  $\tau_{g}$ &  $t_d/t_{c,g}$ \\
\hline
Ratio  of $t_d$ to  $t_{c,h}$ & $\tau_{h}$ &  $t_d/t_{c,h}$ \\
\hline
\end{tabular}
\end{table}
In this section, we simulate the proposed CAJ methods with the EV and NLS methods for the estimation of the JN's channel direction in various scenarios to evaluate the performance of these approaches. In the simulation setup, unless stated otherwise, we used the parameters from Table \ref{tab:param} when a specific parameter is required to be fixed. We set  $N_{t_p}\in \{20, 50\}$,  $N_{t_d}=1000$, and $K \in \{4, 16\}$. We selected $K \in \{4, 16\}$ to illustrate two different scenarios: one where the number of SNs is comparable and close to $K_j$, i.e., $K=4$, and another where $K=16 \gg  K_j$. Additionally, various values of $N_{t_p}$ are used to show the performance of the estimation methods with different numbers of pilot symbols.
 We also define the transmitted SNR for the TN and the JN as $\gamma_{t,s}=P_{t,s}/\sigma^2$ and $\gamma_{t,j}=P_{t,j}/\sigma^2$,  respectively. $\sigma^2$ is the noise variance, and $P_{t,j}$ and $P_{t,s}$ are the transmit power by the JN and the TN, respectively.  For generating instances of jamming and friendly channel vectors, we used a complex normal distribution, i.e., $\mathcal{CN}(0, \textbf{I})$.\par
\subsection{Estimation Performance}
To assess the performance of the estimators, we use the mean squared absolute difference (MSAD) as 
\begin{equation}
   \mathrm{ MSAD}= \mathrm{E}\{\||\widehat{\overline{\textbf{g}}}|-|\overline{\textbf{g}}|\|_2^2\},
\end{equation}
where $|\cdot|$ is the element-wise absolute value of the vector. We use the absolute value because we are specifically interested in the energy difference in each dimension of the vectors, which acts as a metric for the error in the estimation of the direction.\par
\begin{figure}[t!]
    \centering
    \includegraphics[width=3.0in, height=2.35in]{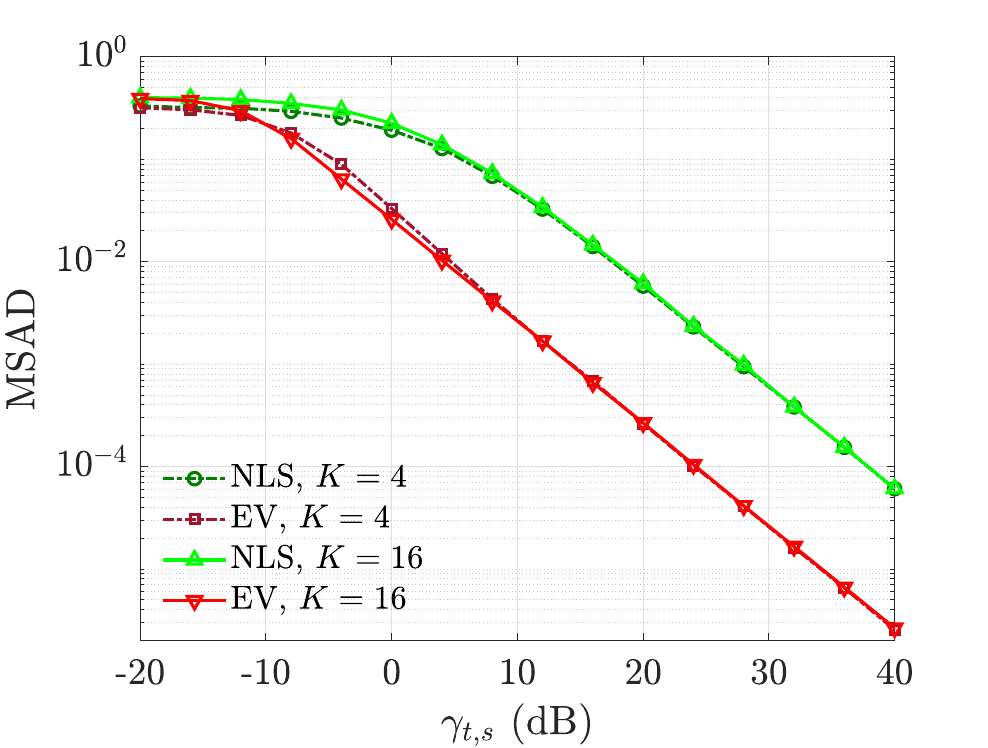}
    \caption{ MSAD of the NLS and  EV  methods versus $\gamma_{t,j}$ for  $N_{t_p}=20$ and $\gamma_{t,s}=10$ dB with $K=4$ and $K=16$.}
    \label{fig:mse_1}
\end{figure}
\begin{figure}[!t]
    \centering
 \includegraphics[width=3.0in, height=2.35in]{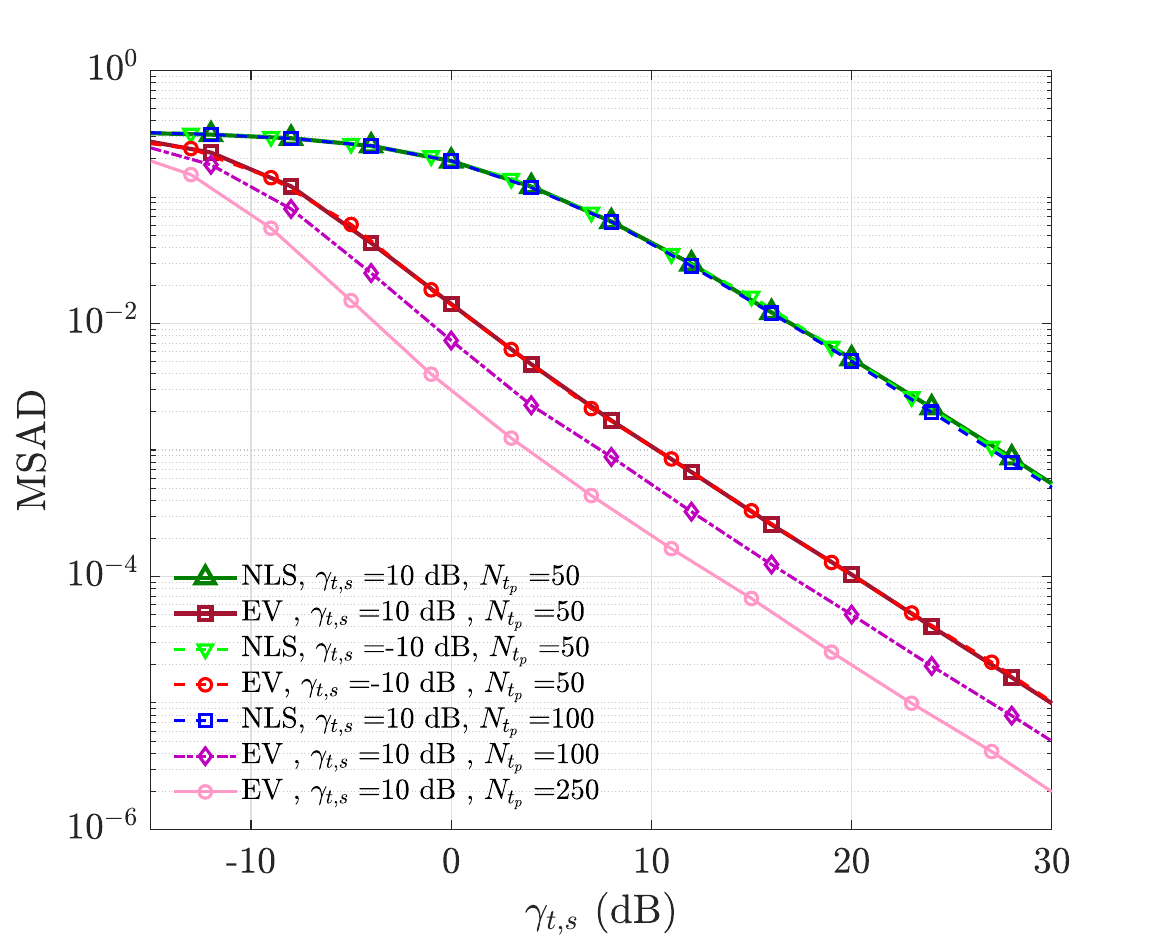}
    \caption{MSAD of the NLS and  EV methods versus $\gamma_{t,j}$ for  $N_{t_p}=50, 100$, and $250$  with $K=4$. }
    \label{fig:mse_2}
\end{figure}
Fig. \ref{fig:mse_1} shows the MSAD for the EV and NLS methods versus the transmit SNRs by the JN for $K=4$ and $16$. The estimation of both methods becomes further accurate when $\gamma_{t,j}$ is larger. The proposed EV method is significantly superior to the NLS. For MSAD equal to $10^{-4}$, the NLS requires about $13$ dB larger $\gamma_{t,j}$. We can also observe that the performance of both methods is quite invariant to  changes in the number of SNs.\par
 Fig. \ref{fig:mse_2} displays similar curves for different numbers of pilot symbols. Only the EV methods show improved performances when increasing the number of pilot symbols. For the EV method, increasing $N_{t_p}$ from $50$ to $250$ yields a higher performance gain. This suggests that one appropriate option to improve the performance of the entire AJ method based on the EV is to increase $N_{t_p}$. In addition, the similarity between the broken lines and solid lines in this figure illustrates that the SNR of the signal transmitted by the TN  has no influence on the performance of the estimation.\par
\begin{figure}[!t]
    \centering
    \includegraphics[width=3.25in, height=2.4in]{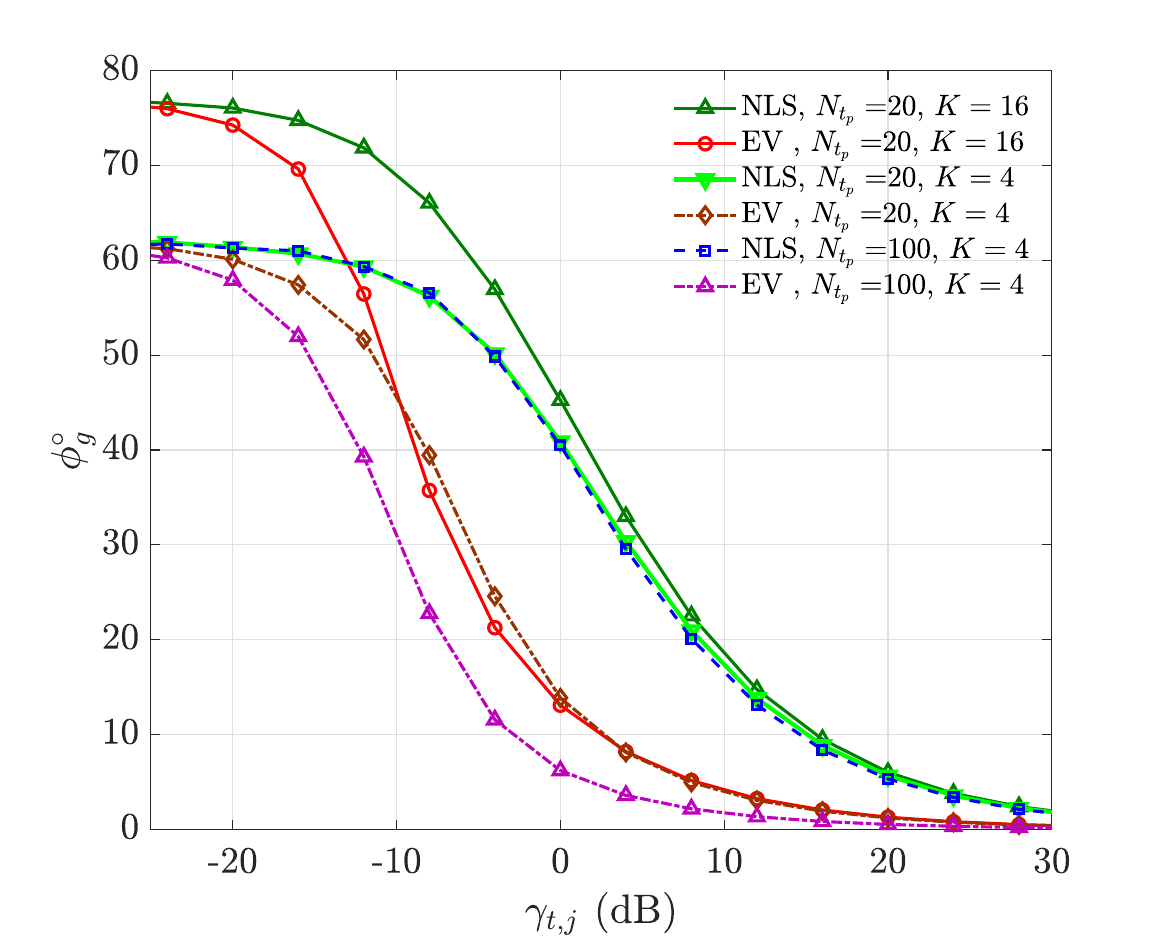}
    \caption{Angular difference between the estimated direction and the true direction versus $\gamma_{t,j}$ with $\gamma_{t,s}=10$ dB, for various values of $K$ and $N_{t_p}$.}
    \label{fig:phi_yj}
\end{figure}
For further analysis, we evaluated the curves using the cosine similarity metric \cite{main_cos_sim}. This metric is obtained as $\mathrm{cos}\phi_g=|\hat{\bar{\textbf{g}}}^H\bar{\textbf{g}}|$, where $\hat{\bar{\textbf{g}}}$ represents the estimated direction of the channel, and $\phi_g$ denotes the angle between the two vectors. When the estimated direction of the jamming channel aligns perfectly with the true direction, $\phi_g$ becomes zero. Fig. \ref{fig:phi_yj} shows $\phi_g$ versus $\gamma_{t,j}$ for both the NLS and EV methods 
for different values of $K$ and $N_{t_p}$. This figure is fully consistent with Fig. \ref{fig:mse_1} and Fig. \ref{fig:mse_2}, where larger $\gamma_{t,j}$ brings the angle closer to zero. We observe that increasing the number of pilot symbols only enhances the performance of the EV method. For example, for $K=4$ and $\gamma_{t,j}=0$ dB, the estimated direction $\hat{\bar{\textbf{g}}}$ by the EV method using $N_{t_p}=100$ and $N_{t_p}=20$ is only about $7^\circ$ and $15^\circ$ apart from the true direction, respectively. However, for the NLS method with $K=4$, $N_{t_p}=100$ or $N_{t_p}=20$, this angle is about $41^\circ$. Additionally, a larger $K$ results in a similar $\phi_g$ at high $\gamma_{t,j}$, while at lower $\gamma_{t,j}$, the performance degrades. This is because, in the low-SNR regime, the estimation method relies on pure noise, making it similar to a blind guess, where increasing dimensionality lowers the chances of a good guess.
\begin{figure}[!t]
    \centering  
    \includegraphics[width=3.25in, height=2.4in]{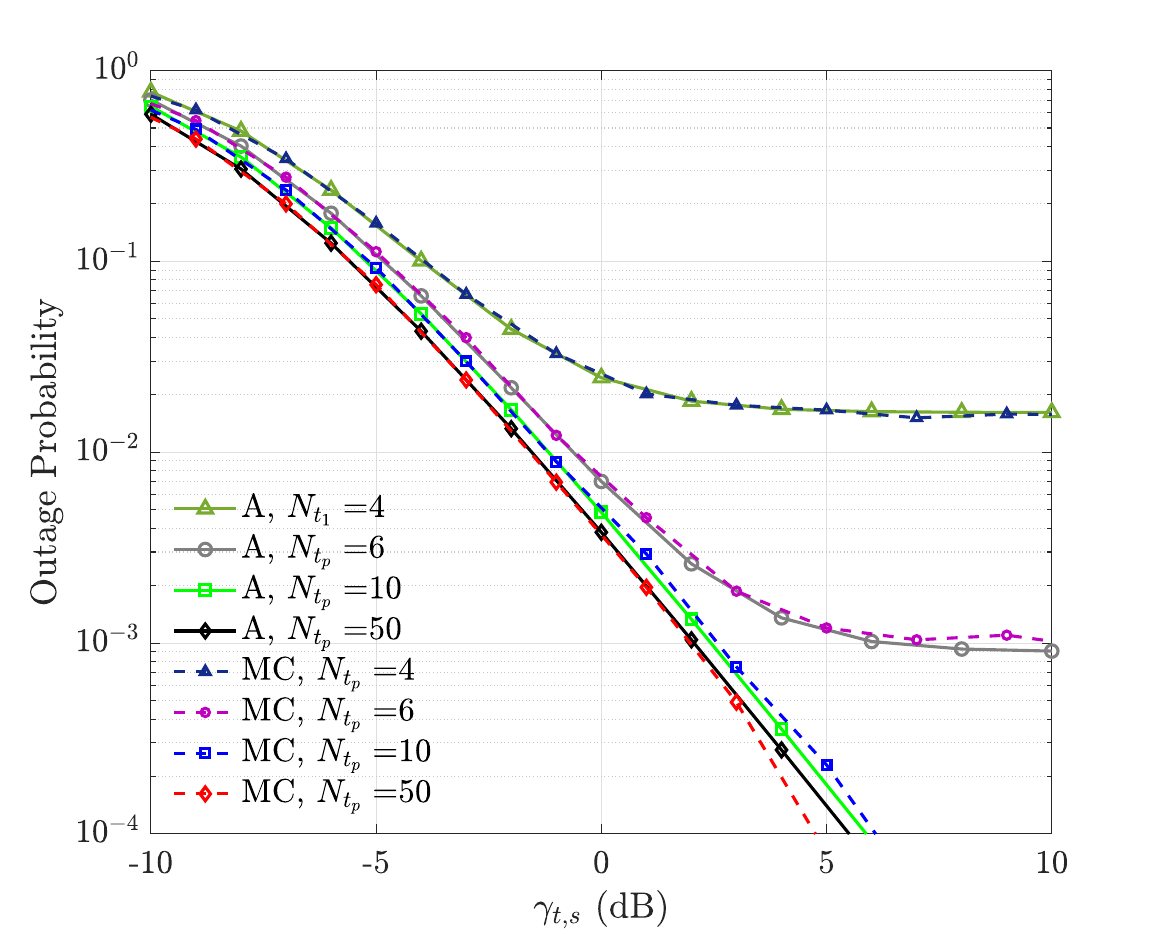}
    \caption{Monte Carlo (MC) and analytical (A) outage probability versus $\gamma_{t,s}$ for the EV-CAJ with $\gamma_{t,j}=40$ dB, $\gamma_{th}=-10$ dB, $N_{t_d}=100$, and $K=4$.}
    \label{fig:po_ysN}
    \end{figure}
\subsection{Outage Probability}
\begin{figure}[!t]
    \centering
\includegraphics[width=3.25in]{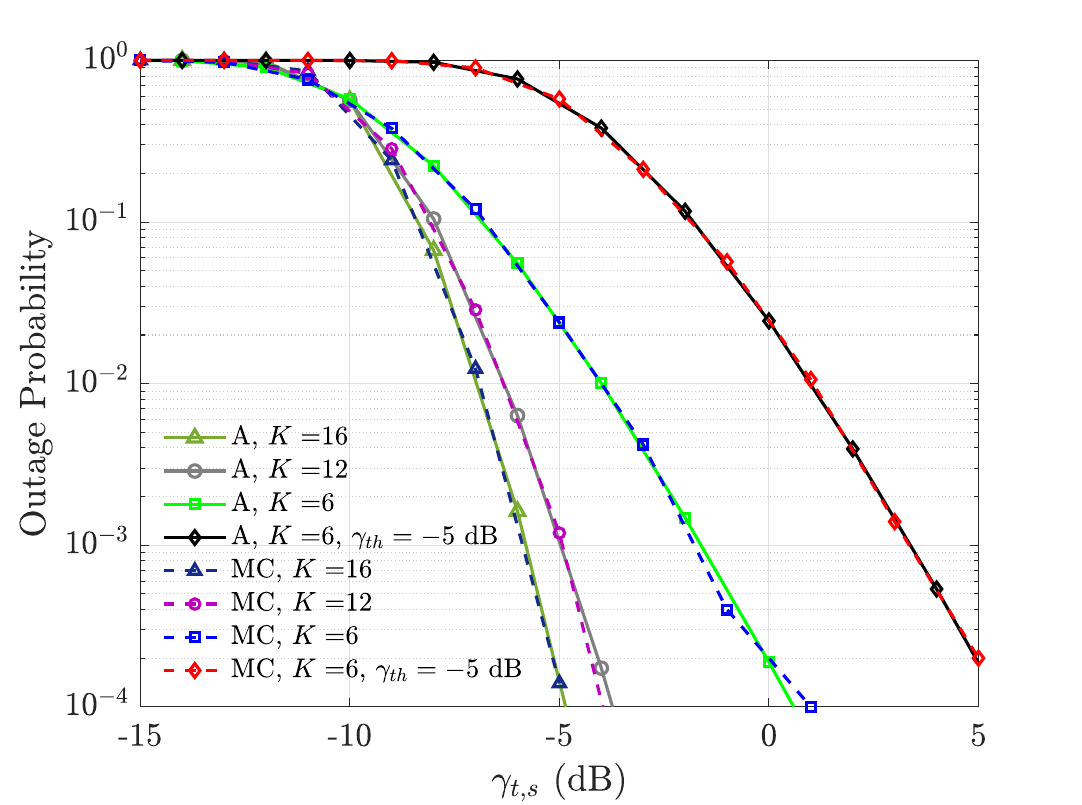}
    \caption{Monte Carlo (MC) and analytical (A) outage probability versus $\gamma_{t,s}$ for the EV-CAJ with $\gamma_{t,j}=40$ dB, $N_{t_d}=100$, and $N_{t_p}=50$. }
    \label{fig:Po_ysK}
\end{figure}

In this subsection, we investigate the joint outage probability performance of the proposed  EV and CAJ methods, namely the EV-CAJ, for various parameters. In addition to MC simulations, labeled with ``MC'', we also plot the analytical curves, labeled with ``A" in Figs. \ref{fig:po_ysN}  and \ref{fig:Po_ysK}.\par
Fig. \ref{fig:po_ysN} shows the outage probability of the EV-CAJ versus the transmitted SNR by the TN, where $K=4$ and $N_{t_d}=100$. The threshold SNR  is adjusted to $\gamma_{th}=-10$ dB. In this analysis, we examine scenarios involving different numbers of pilot symbols. As  $N_{t_p}$ increases, the term 
$\sum_{n=1}^{N_{t_p}} \left| \mathbf{j}_{t_p}^T\boldsymbol{\psi}_n \right|^2$ in equation (\ref{eq27n})
increases. This indicates improved suppression of the jamming signal due to a more accurate estimation of   $\bar{\mathbf{g}}$, resulting in a larger $\beta$, which  reduces $P_{out}$. The same behavior is illustrated in Fig. \ref{fig:po_ysN}. Conversely, when $N_{t_p}$ is small, the estimation error remains high, $\beta$ is constrained, and even increasing $P_{t,s}$ does not further decrease $P_{out}$, resulting in a fixed tail for curves with $N_{t_p}<10$ in Fig. \ref{fig:po_ysN}. To avoid this, it is necessary to have $N_{t_p}\ge 10$ in each time frame.\par
Fig. \ref{fig:Po_ysK} demonstrates the impact of the number of SNs on the outage probability. Since the outage probability of the EV-CAJ can be found by equation (\ref{eq27n}) and CDF of a chi‐square distribution, the outage probability at high SNR is related to 
$P_{out} \propto  (1/\gamma_{t,s} )^K$,
 showing the effect of diversity gain \cite{goldsmith2005wireless}. Therefore, increasing $K$ significantly decreases the outage probability, which is clearly reflected in the steeper decline of $P_{out}$ for large $\gamma_{t,s}$ in Fig. \ref{fig:Po_ysK}. We also consider the case of $\gamma_{th}=-5$ dB, where a higher level of  received SNR is needed to avoid  communication outage. consequently, in this case, the outage probability is considerably higher since a stronger signal is required. The  simulation results for the outage probability align closely  with the analytical derivations presented in Section IV and equation (\ref{eq27n}).
 
\subsection{Symbol Error Probability}
\begin{figure}
    \centering   \includegraphics[width=3.23in, height=2.4in]{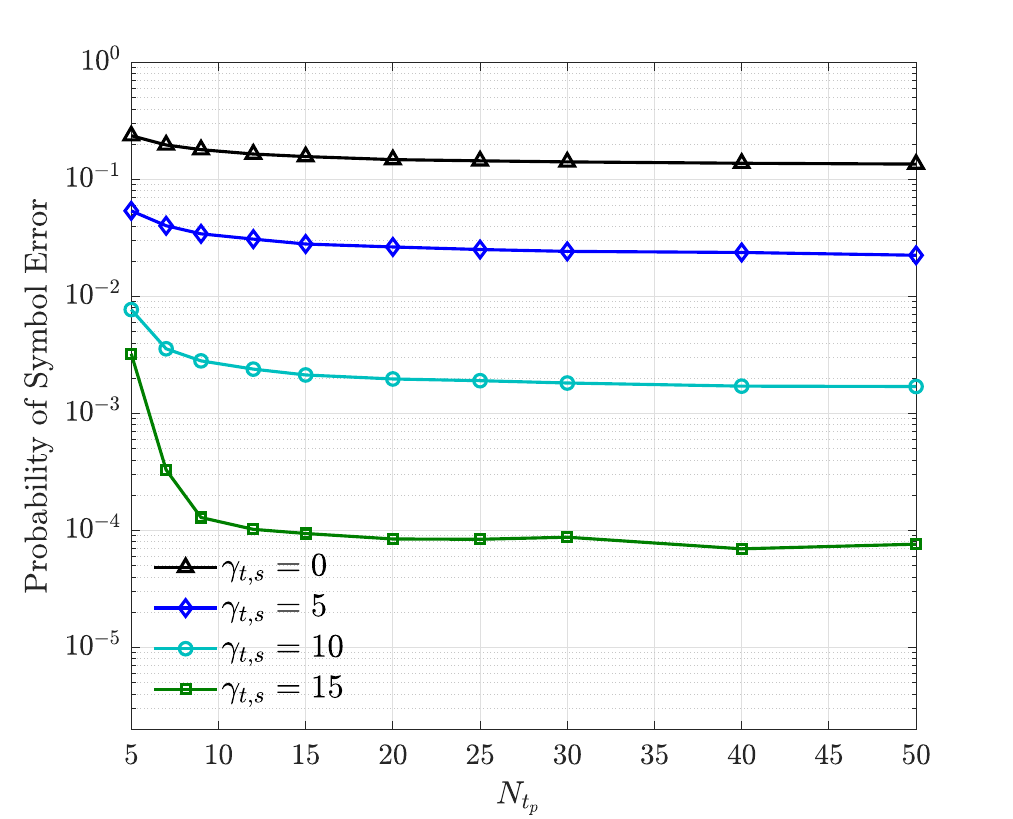}
    \caption{Probability of symbol error versus $N_{t_p}$ for the EV-CAJ method when  $\gamma_{t,j}=40$ dB and $K=4$.}
    \label{fig:Pse_ntp}
\end{figure}
Here, we evaluate the capability of the proposed CAJ method to eliminate the jamming signal generated by the JN. To this end, we simulate the symbol error probability,  $P_{se}$, versus $\gamma_{t,s}$ for the 
case where the JN is transmitting a very strong signal with $\gamma_{t,j}=40$ dB. We generate the transmitted signal by the TN and the JN using QPSK modulation to model a deceptive jamming scenario which is a challenging jamming attack to address \cite{survey_jamming}. \par

As seen in Fig.~\ref{fig:mse_2} and Fig~\ref{fig:phi_yj}, the number of pilot symbols $N_{t_p}$ greatly affects estimation accuracy of the proposed EV method. To further investigate the impact of this estimation error, we simulate $P_{se}$ of the EV-CAJ method for various values of $N_{t_p}$ per time frame, assuming $K = 4$ and $\gamma_{t,j} = 40$~dB in Fig.~\ref{fig:Pse_ntp}. As can be seen in the results, increasing $N_{t_p}$ leads to a notable increase in $P_{se}$, particularly when $N_{t_p} \leq 10$. However, despite a high transmit SNR by the TN, the EV-CAJ method exhibits severely degraded performance at $N_{t_p} = 5$, primarily due to the large estimation error caused by the limited pilot symbols. In this case, the inaccurate estimation of $\bar{\textbf{g}}$ by the proposed EV method results in a poor jamming suppression by the CAJ approach. For $N_{t_p} > 10$, the symbol error probability stabilizes across all values of $\gamma_{t,s}$, suggesting that a further increase of the number of pilot symbols yields a minimal performance improvement. Based on these results, we conclude that setting  $10 < N_{t_P} \leq50$  results in satisfactory estimation and jamming suppression performance of the EV-CAJ method with minimal overhead.\par
\begin{figure}[t!]
    \centering
  \includegraphics[width=3.25in, height=2.4in]{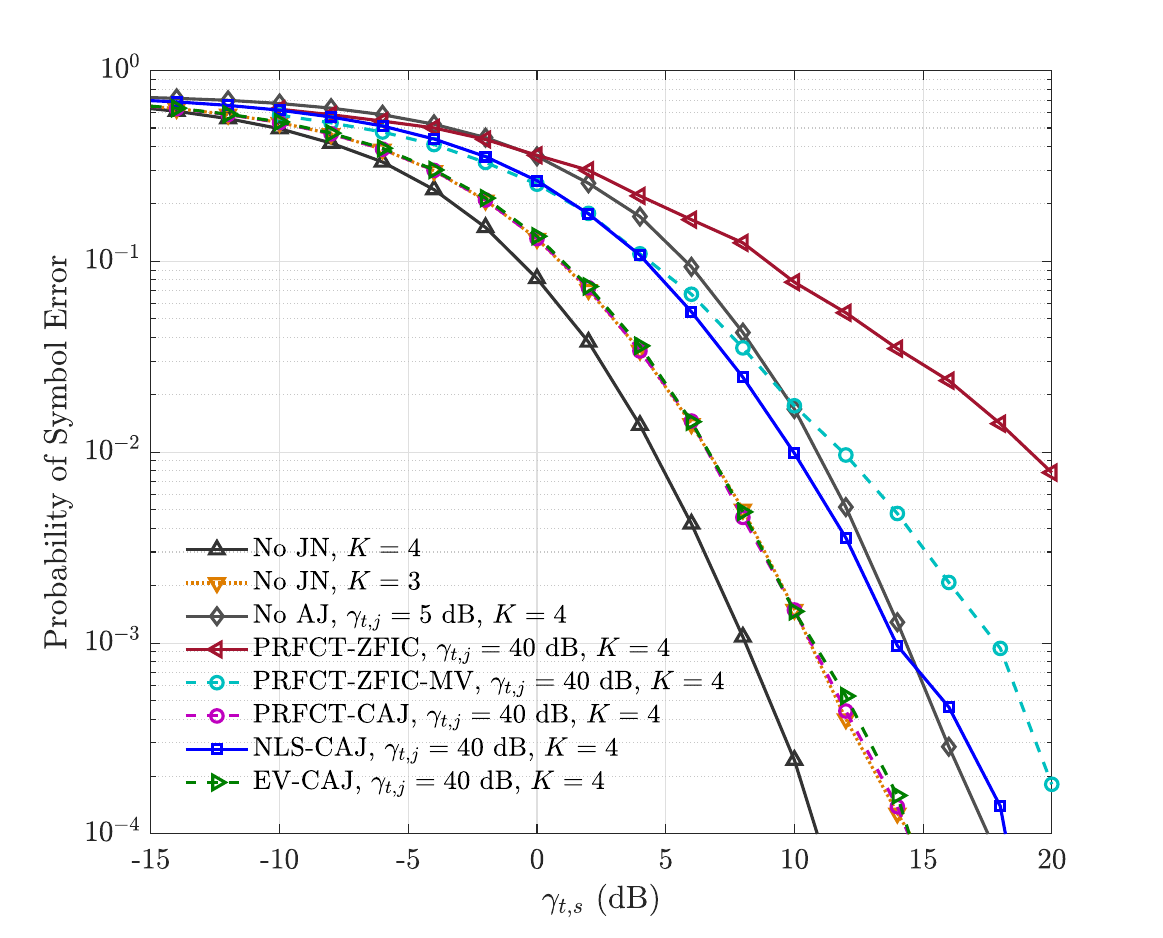}
    \caption{Probability of symbol error versus $\gamma_{t,s}$ for $\gamma_{t,j}=40$ dB, $N_{t_p}=50$, and $K=4$.}
    \label{fig:Pse_y2}
\end{figure}
In the following, we simulate the CAJ scheme with the CSI obtained by the proposed estimation methods. In addition to the EV and NLS estimation methods combined with the CAJ method, we simulate the proposed CAJ strategy with perfect CSI of the jamming channel, namely ``PRFCT-CAJ". To make a more thorough comparison of the suggested CAJ, we also provide the result of the no jamming attack  (``No JN") case, and the case where a mild jamming signal with $\gamma_{t,j}=5$ dB exists with no AJ strategy, namely ``No AJ". In addition, we simulate the zero-forcing interference cancellation (ZFIC) and ZIFC with majority vote (ZIFC-MV) methods, proposed in \cite{ZFIC,ZFIC-conf}. To simulate these methods, we use the perfect CSI of the JN's channel. Thus, the two methods are referred to as ``PRFCT-ZFIC" and ``PRFCT-ZFIC-MV". 
\par
Fig. \ref{fig:Pse_y2} shows the simulated $P_{se}$ versus $\gamma_{t,s}$ for $K=4$ and  $N_{t_p}=50$. We observe that the proposed EV-CAJ attains $P_{es}=10^{-3}$ with  $\gamma_{t,s}$  roughly  $2.5$ dB higher than the No JN case.  The NLS-CAJ requires an SNR about $7$ dB larger than the case where no JN is present to attain the same $P_{es}$. It is also observed that methods based on ZFIC  show inferior performances even though perfect CSI of the JN's channel is used. This figure also shows that the EV method achieves a very accurate estimation of the jamming channel since the performance of the EV-CAJ is very close to that of the CAJ with perfect CSI on the JN's channel. In addition, the curve of the EV-CAJ method with $K=4$ is very similar to that of the No JN case with $K=3$. The reason behind this is that we removed the jamming with the use of $\textbf{G}^\perp$, which transforms the received signal into a new space with $K-1$ dimensions. This new space is the orthogonal complement space of the line space spanned by $\textbf{g}$. Therefore, we can effectively eliminate a large portion of the jamming signal at the cost of losing one DoF.\par
\begin{figure}[!t]
    \centering
    \includegraphics[width=3.25in, height=2.4in]{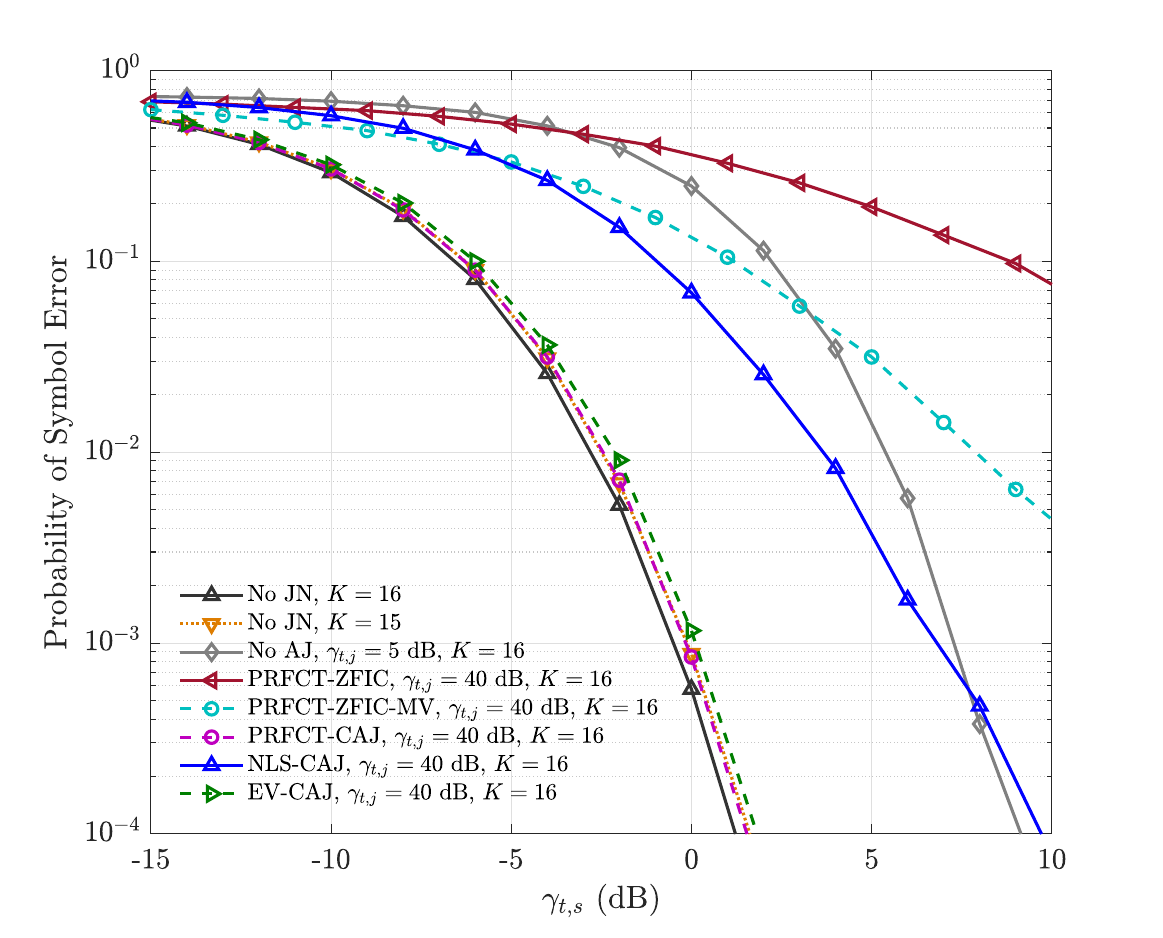}
    \caption{Probability of symbol error versus $\gamma_{t,s}$ for $\gamma_{t,j}=40$ dB, $N_{t_p}=20$, and $K=16$.}
    \label{fig:Pse_y3}
\end{figure}

Increasing the number of SNs significantly improves the performance of the CAJ methods as shown in Fig. \ref{fig:Pse_y3}. Specifically, when there is no jamming attack, $P_{se}=10^{-3}$ is achieved with $\gamma_{t,s}=-0.5$ dB, while the EV-CAJ method can achieve the same $P_{se}$ with $\gamma_{t,s}=0.2$ dB under a severe jamming attack with $\gamma_{t,j}=40$ dB. In this figure, we set $N_{t_p}=20$ and observe that the gap between the PRFCT-CAJ and the EV-CAJ is larger compared to Figure \ref{fig:Pse_y2}. This increased gap is due to the smaller number of pilot symbols, which results in a less accurate estimation of the channel direction for the JN. The results of this simulation scenario with a larger number of DoFs show the excellent performance of the EV-CAJ method since it only requires an increase of $0.7$ dB to the SNR to overcome the strong deceptive jamming attack.\par
We observe that increasing 
$K$ and the size of the signal space helps the CAJ method perform more effectively due to the increased  DoFs. These additional DoFs improve the system’s ability to separate and remove the jamming signal without damaging the desired signal. Thus, adding more SNs to the network enables the FC to effectively counter severe jamming attacks. However, increasing the number of SNs in commercial networks requires additional hardware, which might be expensive and impractical for many commercial applications. Furthermore, the simulation results illustrate that the proposed methods are influenced by the SNR from both the TN and JN, which depends on the distance between transmitters and the SNs. However, it is important to note that the direction of arrival (DOA) of the received signals has no impact on the performance of the proposed methods, as we did not make any assumptions about DOA nor used it in deriving the proposed methods.

\subsection{Multiple JNs and TNs}
\begin{figure}[!t]
    \centering
    \includegraphics[width=3in]{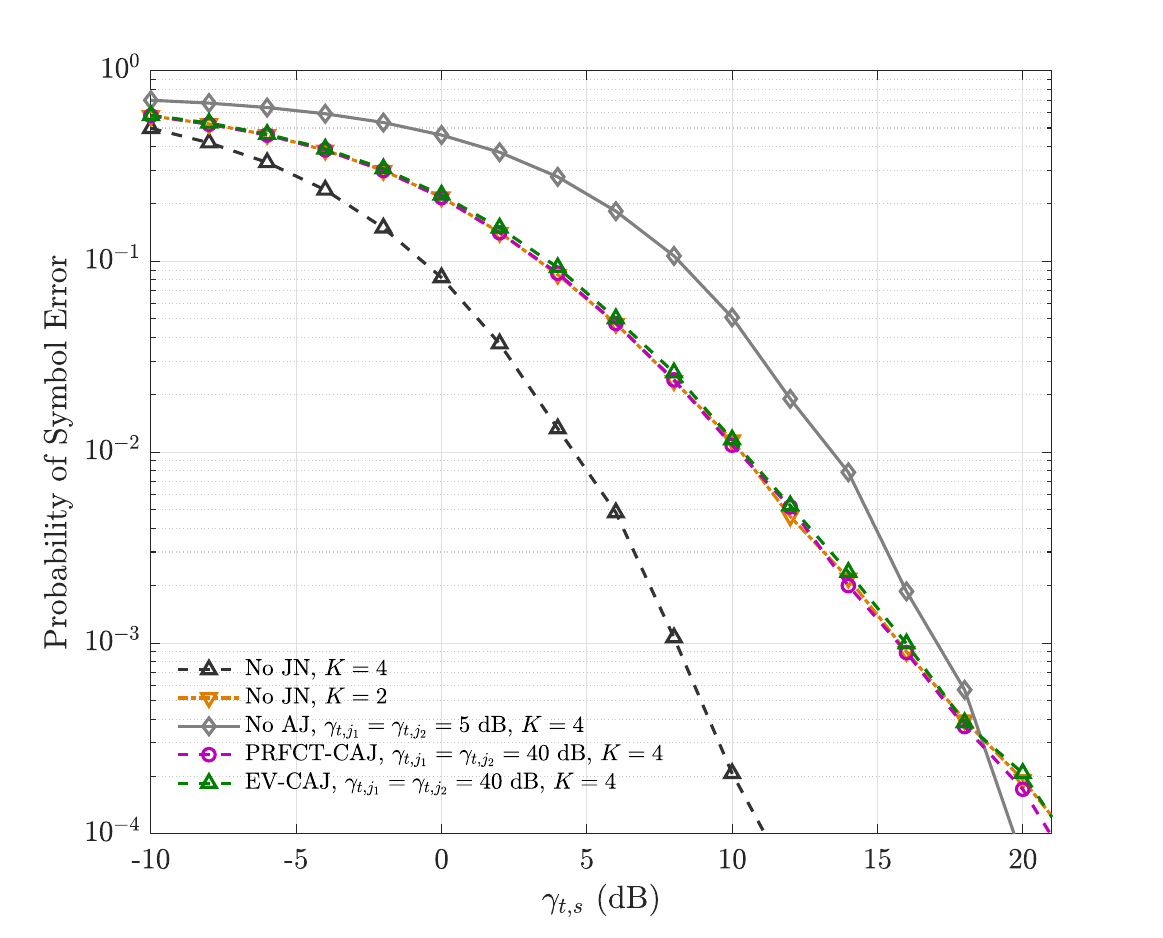}
    \caption{Probability of symbol error versus $\gamma_{t,s}$ for $\gamma_{t,j_1}=\gamma_{t,j_2}=40$ dB, $N_{t_p}=50$, and $K=4$.}
    \label{fig:Pse_y_2JN_k4}
\end{figure}
One important capability of the proposed EV-CAJ method is that it can also be applied to combat $K_j$ malicious JNs, as discussed in Section \ref{sec_multiple_caj}. We noted that as long as $K-K_j>0$, we can mitigate the JNs' impact on friendly communications. Obviously, with a higher remaining DoF, i.e., $K-K_j$, the performance of  EV-CAJ  becomes closer to that of the No JN case.
Here, we consider $K_j=2$ deceptive JNs attacking friendly communications. We set the transmitted power of JNs to  $\gamma_{t,j_1}=\gamma_{t,j_2}=40$ dB, where  $\gamma_{t,j_1}$ and $\gamma_{t,j_2}$ show the transmit SNR by JN1 and JN2, respectively. In addition, we simulate the No JN scenarios with $K$ and $K-K_j$ SNs and the No AJ strategy with two JNs with a transmit SNRs equal to $\gamma_{t,j_1}=\gamma_{t,j_2}=5$ dB for thorough comparisons with the suggested EV-CAJ approach.\par
In Fig. \ref{fig:Pse_y_2JN_k4}, the number of SNs is set $K=4$, and the number of pilot symbols is $N_{t_p}=50$. We observe that the proposed EV-CAJ, the PRFCT-CAJ, and No JN with $K=2$ show similar results. This observation confirms that the EV-CAJ can reach the performance of the ideal case with no jamming and $K-K_j$ DoFs, where increasing the number of JNs only causes $K-K_j$ becomes smaller for the EV-CAJ. For example, in this figure, having two JNs is equivalent to losing two SNs. To compensate for this loss and achieve $P_{se} = 10^{-2}$,  the EV-CAJ requires $\gamma_{t,s}=10$ dB which is roughly $5.5$ dB greater than the required SNR for the No JN case to achieve the same $P_{se}$. \par
\begin{figure}
    \centering
    \includegraphics[width=3in]{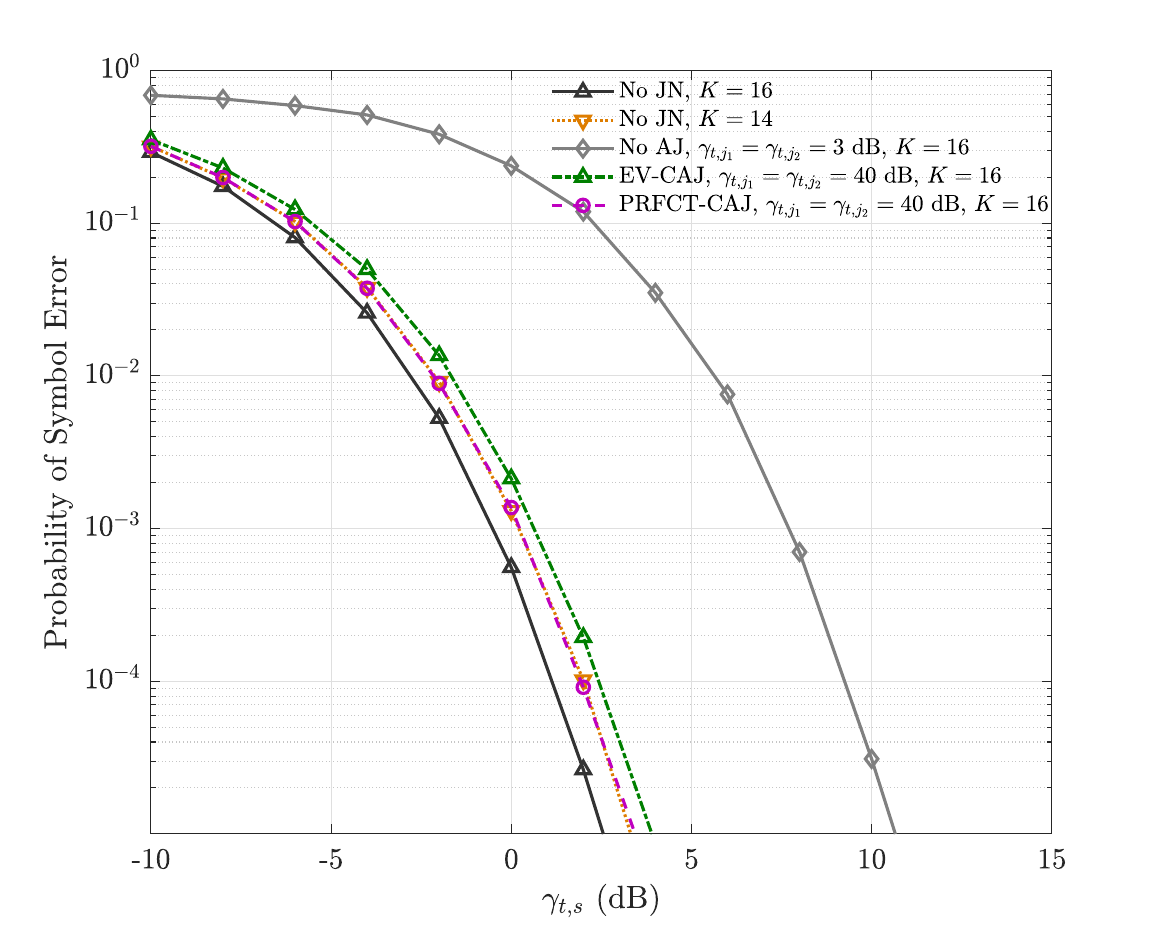}
    \caption{Probability of symbol error versus $\gamma_{t,s}$ for $\gamma_{t,j_1}=\gamma_{t,j_2}=40$ dB, $N_{t_p}=20$, and $K=16$.}
    \label{fig:Pse_y_2JN_K16}
\end{figure}

 
Fortunately, the effect of multiple JNs can be neutralized by increasing $K$. Fig. \ref{fig:Pse_y_2JN_K16} shows the same results for $K=16$ and $N_{t_p}=20$. The EV-CAJ method needs just about a $1.2$ dB greater $\gamma_{t,s}$ than the No JN case with $K=16$ to attain $P_{se}=10^{-3}$. Additionally, we can observe that the EV-CAJ method with approximately $0.4$ dB higher SNR achieves the same performance as the PRFCT-CAJ and No JN cases with $K=14$. This difference between the performance of the EV-CAJ and the PRFCT-CAJ methods can be attributed to the small number of pilot symbols used in the EV-CAJ for the estimation of the space spanned by vector channels of  JN1 and JN2 in comparison to Fig. \ref{fig:Pse_y_2JN_k4}. It is observed that, with a larger
number of SNs, the performance of the EV-CAJ method is just $1.2$ dB inferior to the case of no jamming attack which highlights its outstanding capability in combating multiple sources of deceptive jamming signals with a very high power. On the other hand, Fig.  \ref{fig:Pse_y_2JN_K16}  shows that in the absence of an appropriate AJ method, a jamming attack by two JNs with  $\gamma_{t,j_1}=\gamma_{t,j_2}=3$  dB, severely increases the probability of error.\par
\begin{figure}[!t]
    \centering
    \includegraphics[width=3in]{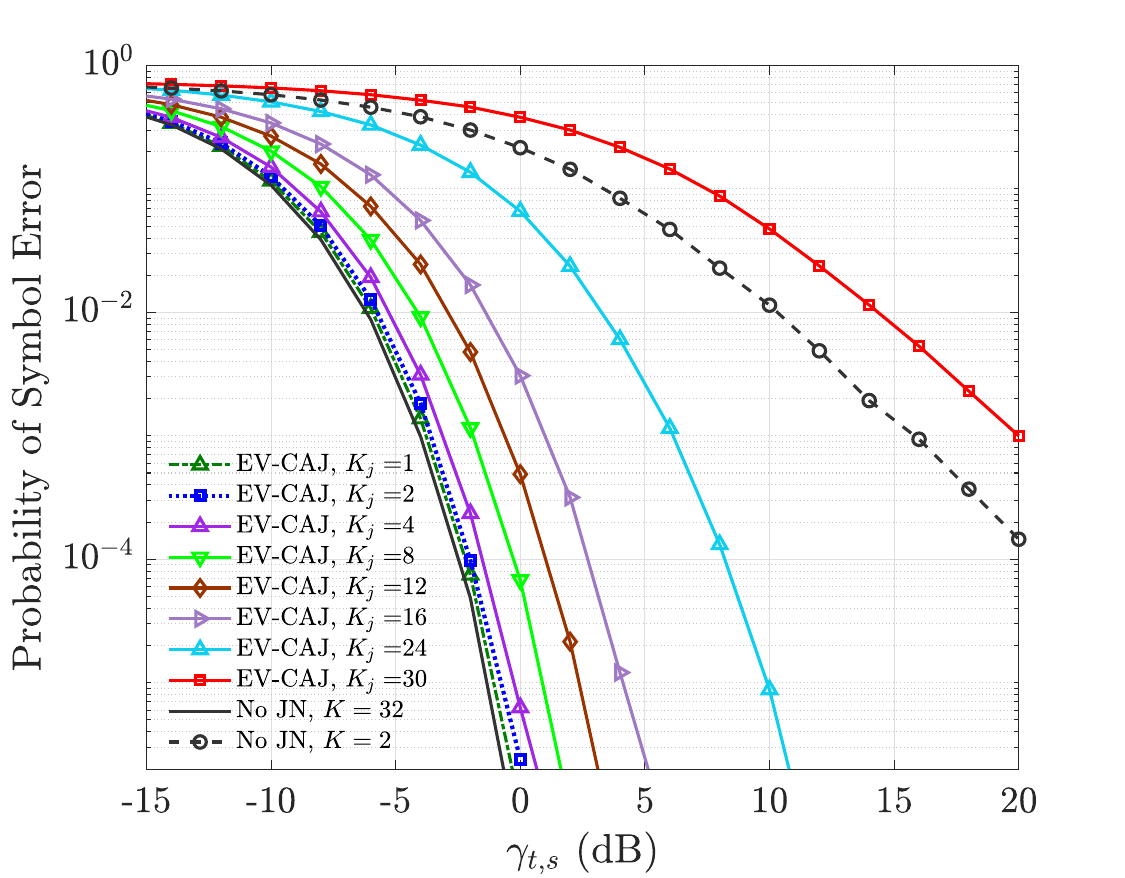}
    \caption{Probability of symbol error versus $\gamma_{t,s}$ for different $K_j$, when $\gamma_{t,j} = 40$ dB, $N_{t_p} = 50$, and $K = 32$.}
    \label{fig:MJ32}
\end{figure}
 Fig. \ref{fig:MJ32} illustrates the  probability of symbol error versus $\gamma_{t,j}$ for the EV-CAJ method with $K=32$ and $N_{t_p}=50$, across different $K_j$ values.  The transmit SNR by  JNs is set to $\gamma_{j,t}=40$ dB for all  JNs. The results indicate that EV-CAJ's performance approaches the No JN scenario when fewer JNs are active, resulting in a larger $K - K_j$. Specifically, for $K_j \leq 4$, EV-CAJ's performance is less than 2 dB inferior compared to the No JN case. Furthermore, at $P_{se} = 10^{-2}$, with $K_j = 30$ (i.e., $K - K_j = 2$), the performance degradation is about 4 dB relative to the No JN scenario with $K = 2$. This demonstrates that EV-CAJ can effectively counteract jamming attacks from a substantial number of JNs.\par
\begin{figure}
    \centering
    \includegraphics[width=3.0in, height=2.35in]{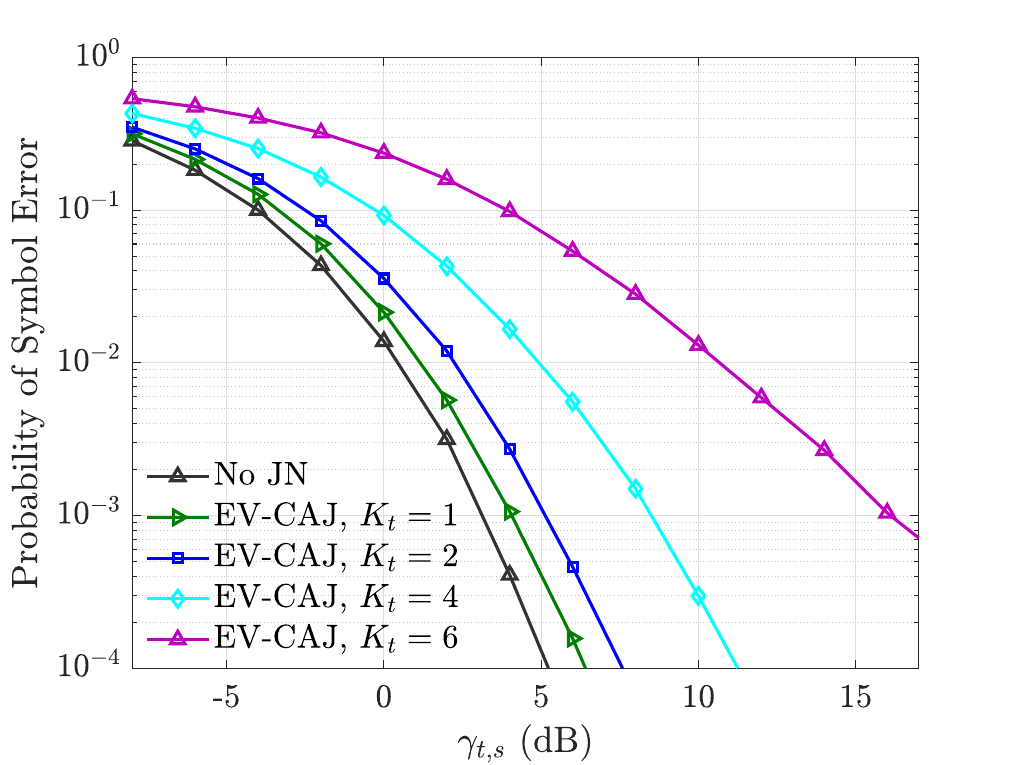}
    \caption{Probability of symbol error versus $\gamma_{t,s}$ for various $K_t$ with $\gamma_{t,j}=40$ dB, $N_{t_p}=50$, and $K=8$.}
    \label{fig:MultiTNs}
\end{figure}
In  Fig. \ref{fig:MultiTNs}, we analyze a network with $K=8$ SNs communicating with different numbers of $K_t$, while being jammed by one JN ($K_j=1$) with a transmit SNR of $\gamma_{t,j}=40$ dB. The number of pilot symbols is set to $N_{t_p}=50$. Each curve in the figure illustrates the symbol error probability, $P_{se}$, for one TN. For a single TN using the EV-CAJ, an SNR of $\gamma_{t,s}=4$ dB is needed to achieve $P_{se}=10^{-3}$, which is only about 1 dB higher than in the absence of a JN.  As the number of TNs increases, the network can support more data streams, though at the cost of a higher $P_{se}$. For example, with $K_t=2$ and $K_t=4$, each user must transmit at around $\gamma_{t,s}=5$ dB and $\gamma_{t,s}=8.5$ dB, respectively, to maintain $P_{se}=10^{-3}$ in the FC after jamming neutralization by the EV-CAJ method. This simulation result illustrates that EV-CAJ successfully  supports multiple TNs and users during a jamming attack, adapting to increased demands at the expense of a higher $P_{se}$. 
\subsection{CAJ Under Time-Varying Channels}
The JN can be a fixed node with a slow-fading channel. Alternatively, it can be a mobile source of jamming signals, like a drone or a car with high speeds. In this case, the channel between the JN and SNs would follow a fast-fading model, which makes the jamming attack more challenging. Thus, it is important to discuss the capacity of AJ strategies to handle such types of JNs. \par
The proposed CAJ method in this paper is derived based on the assumption that the channel gains are time-invariant (TI) during the time frame, as shown in Fig. \ref{fig0}. In this section, we aim to examine the effectiveness of the proposed CAJ under the opposing scenario where the channels are fast-fading. Thus, here, we assume that fast-fading channels follow the model developed by Jake \cite{goldsmith2005wireless}. Accordingly, the autocorrelation function of samples of discrete channels follows the Bessel function of zeroth order $R_h(\Delta n)= J_0(2\pi f_{d,h} \Delta nt_s)$ during the interval $t_d$  while the TN transmits its data. Here, $f_{d,h}$ is the Doppler frequency of the channel between the TN and an SN, which is related to the coherence time of the channel as $t_{c,h}=0.423/f_{d,h}$ \cite{rappaport2010wireless}. If we denote the symbol duration by $t_s$,  we have  $t_s= t_d/N_{t_d}$.
 The proposed CAJ method applies the EV or NLS to estimate $\overline{\textbf{g}}$, and then, estimates  $\textbf{h}$ during $t_p$ to detect the desired signal transmitted during $t_d$. Thus, the method is heavily dependent on the CSI of both $\textbf{g}$ and \textbf{h}. Since we assume that the channel vectors are fast fading in this subsection, we denote them by $\textbf{h}[n]$ and $\textbf{g}[n]$.
Here, for the autocorrelation function of $\textbf{g}[n]$, we have  $R_g(\Delta n)= J_0(0.846\pi \tau_g \Delta n/N_{t_d})$. The parameter $\tau_g$ is the ratio of the data transmission interval to the coherence time of the channel. For the JN's channel, $\tau_g=t_d/t_{c,g}$, where we assume that the coherence time of all elements of $\textbf{g}[n]$ is identical and denoted by $t_{c,g}$. A similar assumption and autocorrelation are also applied to the TN's channel vector. As the wireless environment changes rapidly, the coherence time of the wireless channel is likely to become shorter. However, depending on the dynamism of the wireless environment, the time interval between the two consecutive pilot blocks ($t_d$) can be reduced in order to have a smaller data block compared to the coherence time of the channel. It is suggested that a shorter $t_d$ can enhance the robustness of the communication against  fast fading at the cost of reducing the data rate.\par
\begin{figure}[t]
    \centering
    \includegraphics[width=3.0in, height=2.35in]{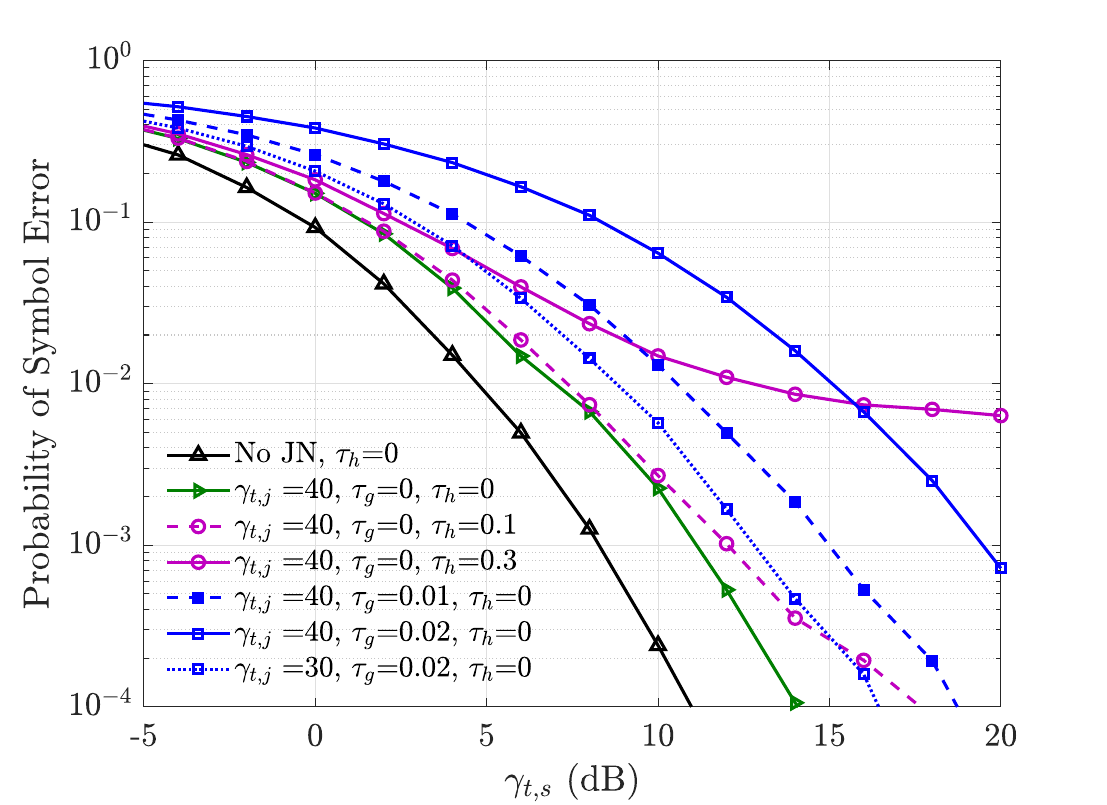}
    \caption{Probability of symbol error versus $\gamma_{t,s}$ for $\gamma_{t,j_1}=\gamma_{t,j_2}=40$ dB, $N_{t_p}=20$, $K=4$, and various values of $\tau_g$ and $\tau_h$.}
    \label{fig:Pse_y_K4_tv}
\end{figure}
\begin{figure*}
    \centering
    \begin{subfigure}[b]{0.20\textwidth}
         \centering
     \includegraphics[width=3.8cm,height=3cm]{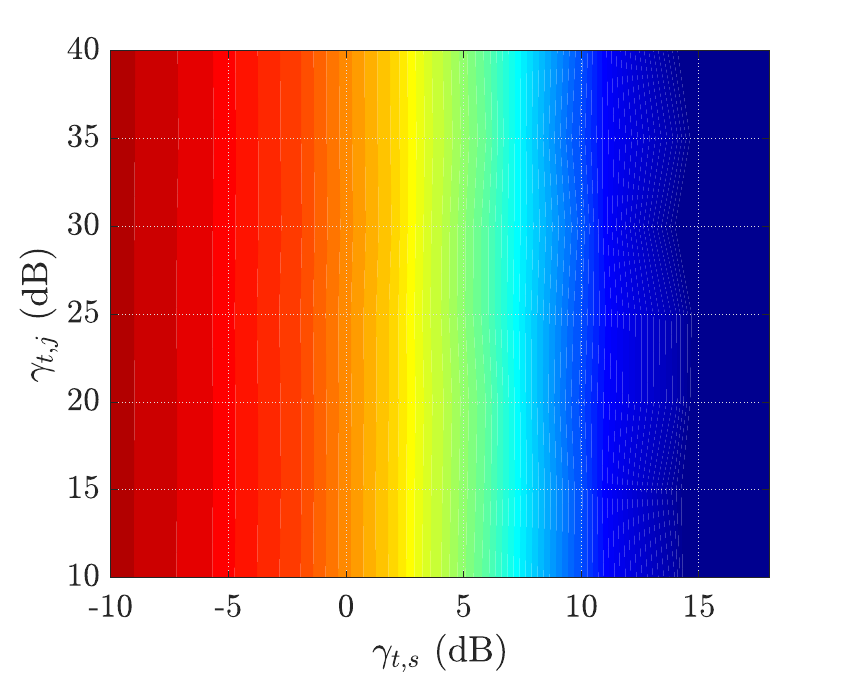}
         \caption{$\tau_g=0$}
         \label{fig:five over x}
     \end{subfigure}
    \begin{subfigure}[b]{0.19\textwidth}
         \centering
        \includegraphics[width=3.8cm,height=3cm]{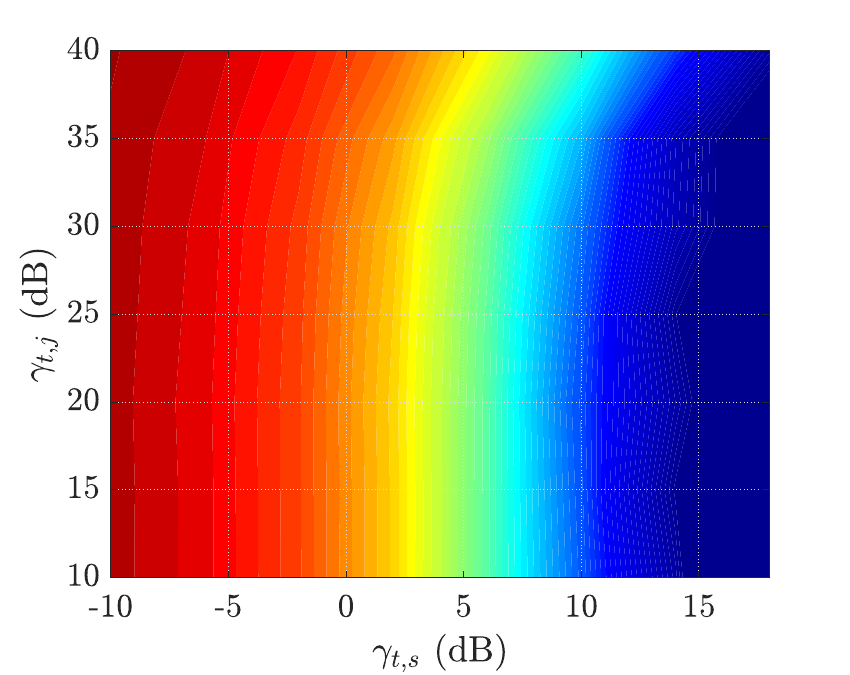}
         \caption{$\tau_g=0.01$}
         \label{fig:five over x}
     \end{subfigure}
     \begin{subfigure}[b]{0.19\textwidth}
         \centering
         \includegraphics[width=3.8cm,height=3cm]{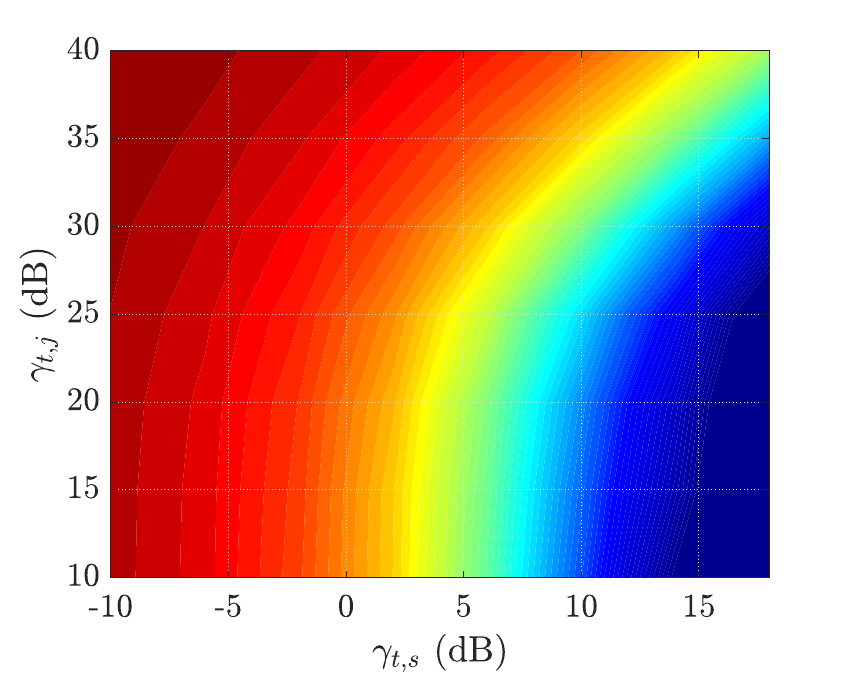}
         \caption{$\tau_g=0.04$}
         \label{fig:five over x}
     \end{subfigure}
     \begin{subfigure}[b]{0.19\textwidth}
         \centering
         \includegraphics[width=3.8cm,height=3cm]{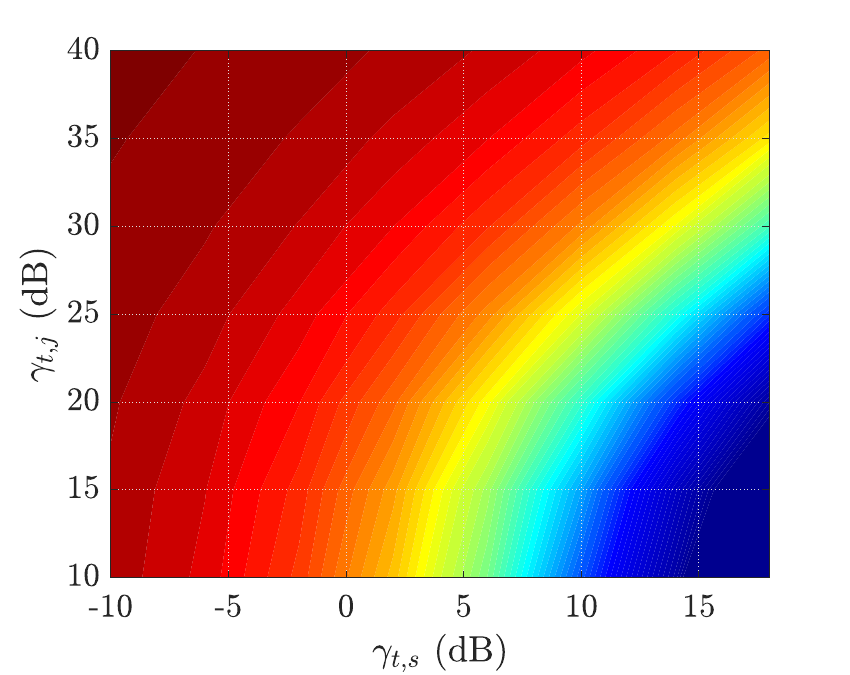}
         \caption{$\tau_g=0.1$}
         \label{fig:five over x}
     \end{subfigure}
     \begin{subfigure}[b]{0.2\textwidth}
         \centering
    \includegraphics[width=4.1cm,height=3cm]{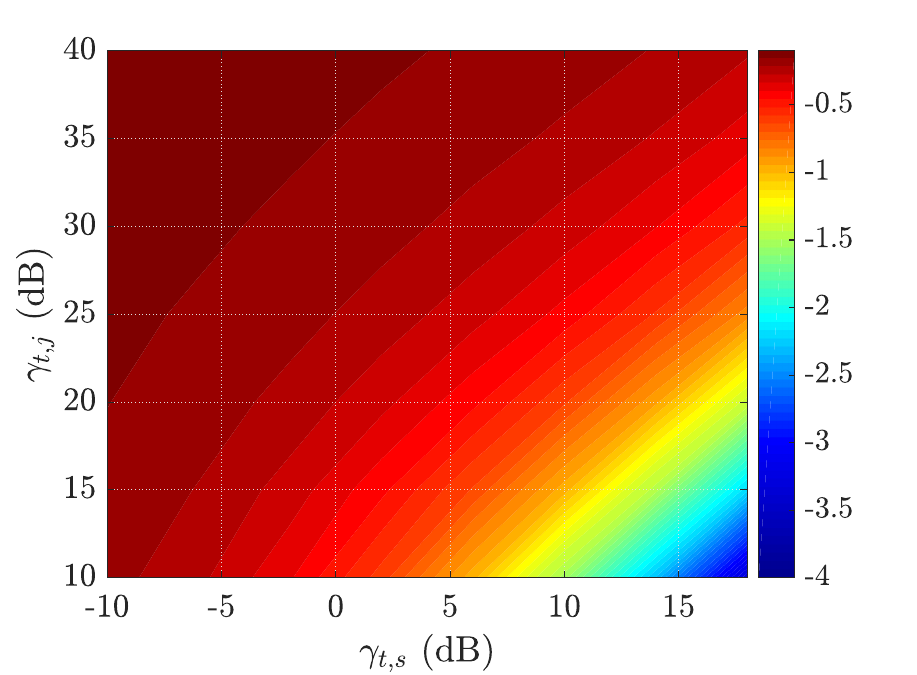}
         \caption{$\tau_g=0.5$}
         \label{fig:five over x}
     \end{subfigure}
     \caption{$P_{se}$ versus $\gamma_{t,j}$ and $\gamma_{t,s}$ for the EV-CAJ scheme with  $\tau_h=0$, $K=4$, and $N_{t_p}=20$. Each subfigure shows a different $\tau_g$.} \label{tv_3d_k4}
\end{figure*}
\begin{figure*}
    \centering
    \begin{subfigure}[b]{0.20\textwidth}
         \centering
     \includegraphics[width=3.8cm,height=3cm]{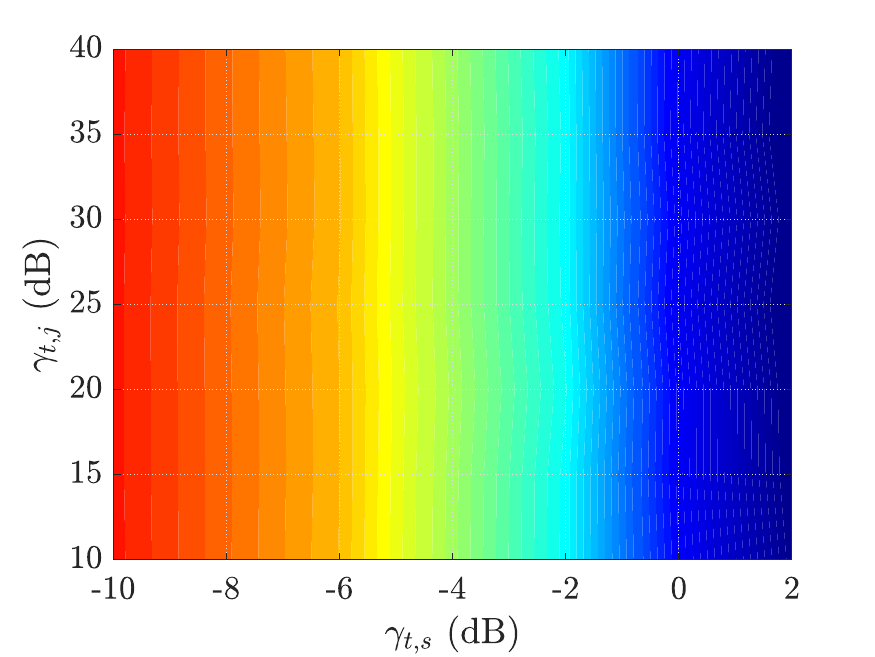}
         \caption{$\tau_g=0$}
         \label{fig:tv_16_a}
     \end{subfigure}
    \begin{subfigure}[b]{0.19\textwidth}
         \centering
         \includegraphics[width=3.8cm,height=3cm]{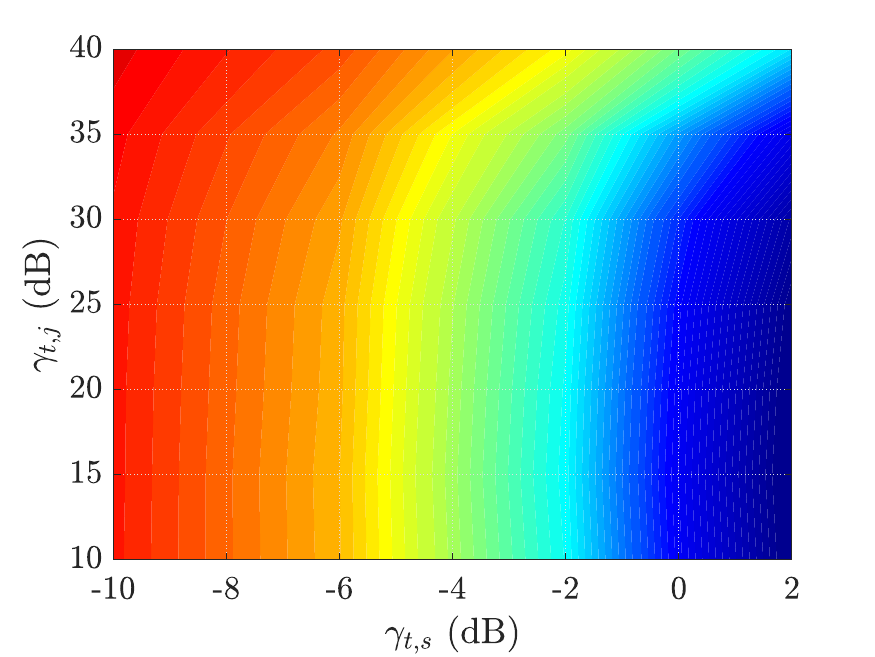}
         \caption{$\tau_g=0.01$}
         \label{fig:five over x}
     \end{subfigure}
     \begin{subfigure}[b]{0.19\textwidth}
         \centering
         \includegraphics[width=3.8cm,height=3cm]{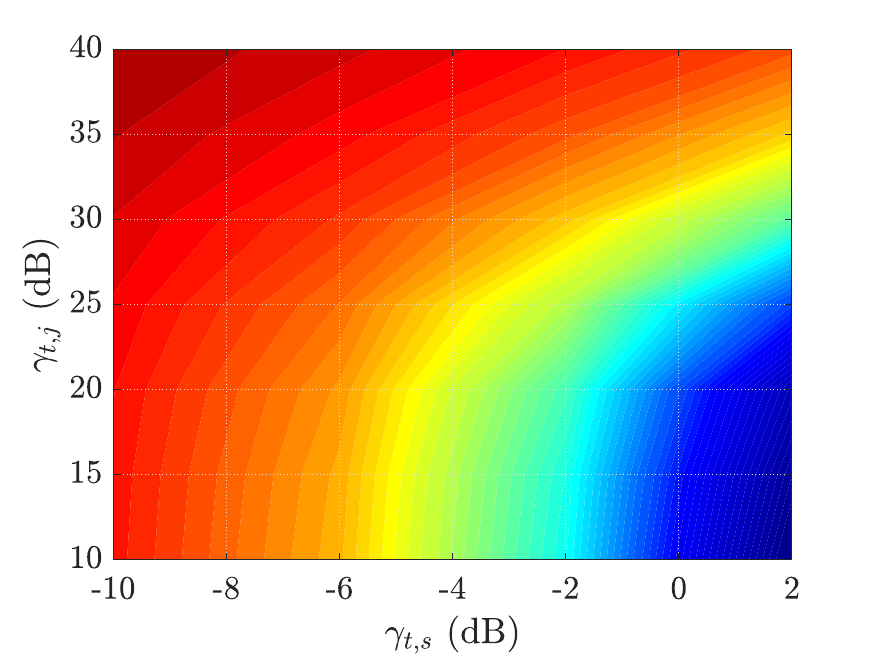}
         \caption{$\tau_g=0.04$}
         \label{fig:five over x}
     \end{subfigure}
     \begin{subfigure}[b]{0.19\textwidth}
         \centering
         \includegraphics[width=3.8cm,height=3cm]{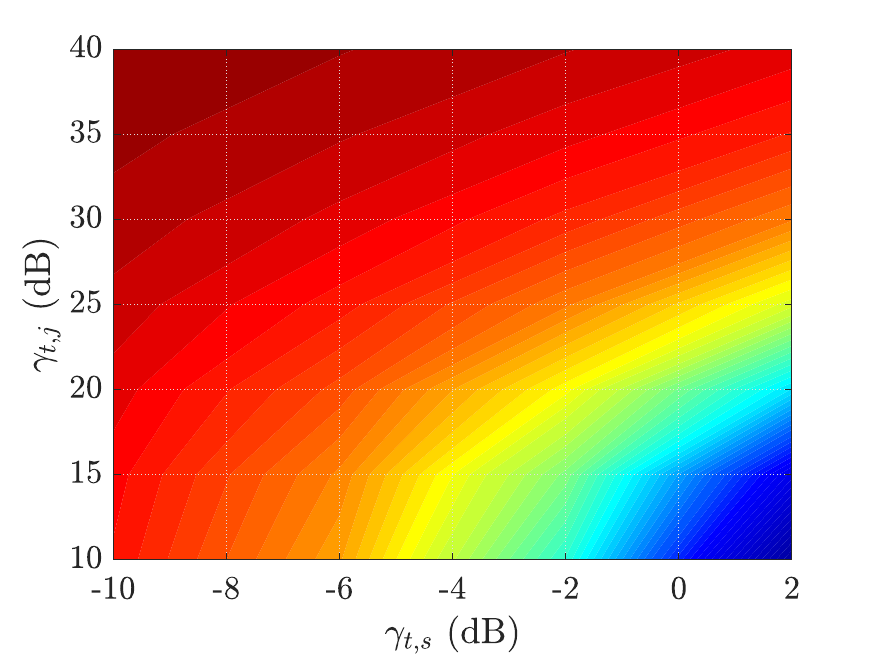}
         \caption{$\tau_g=0.1$}
         \label{fig:five over x}
     \end{subfigure}
     \begin{subfigure}[b]{0.2\textwidth}
         \centering
    \includegraphics[width=4.1cm,height=3cm]{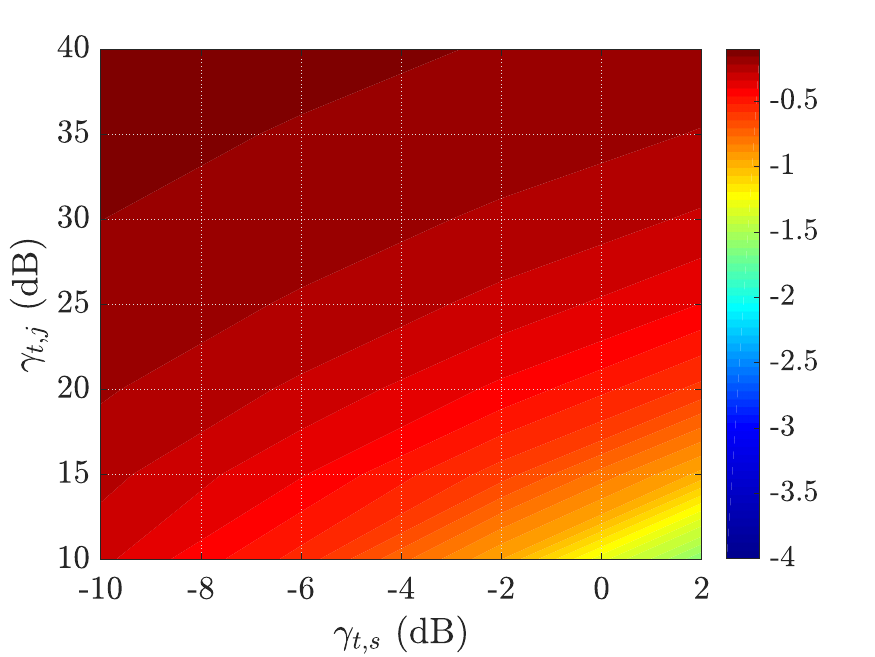}
         \caption{$\tau_g=0.5$}
         \label{fig:five over x}
     \end{subfigure}
     \caption{$P_{se}$ versus $\gamma_{t,j}$ and $\gamma_{t,s}$ for the EV-CAJ scheme with $\tau_h=0$, $K=16$, and $N_{t_p}=20$. Each subfigure shows a different $\tau_g$. }\label{tv_3d_k16}
\end{figure*}
 Fig. \ref{fig:Pse_y_K4_tv} illustrates $P_{se}$ of the EV-CAJ with different channel conditions. For the TV channel between SNs and the TN,  denoted by $\textbf{h}[n]$, with $\tau_h=0.1$, the performance of the EV-CAJ is close to that of the fixed channel. This implies that the EV-CAJ is robust against fast-fading $\textbf{h}[n]$ with $\tau_h\le0.1$. However, when the value of  $\tau_h$ is set to $0.3$, the tail of  $P_{se}$ remains fixed for larger values of $\gamma_{t,s}$. In this case, increasing the transmit power is ineffective in reducing the error rate since the channel varies very rapidly compared to $t_d$ and causes misdetection in the receiver. It is observed that a fast-fading $\textbf{g}[n]$ results in higher $P_{se}$ in comparison with the fixed channel. For instance, to achieve $P_{se}=10^{-3}$ under jamming with $\gamma_{t,j}=40$ dB, the EV-CAJ requires $1.7$ dB  and $8.4$ dB stronger SNRs for $\tau_g=0.01$ and $\tau_g=0.02$, respectively, compared to the fixed $\textbf{g}$ case. However, under a weaker jamming signal with $\gamma_{t,j}=30$ dB and  $\tau_g=0.02$, the degradation is noticeably reduced.\par
 To further investigate the joint effect of $\gamma_{t,j}$ and $\tau_g$, Figs. \ref{tv_3d_k4} and \ref{tv_3d_k16} are presented. These two figures contain contours of $P_{se}$ attained by the EV-CAJ method under various $\tau_g$ and different transmit SNRs by the  JN and the TN.  Both figures include five subfigures where each subfigure presents a $\textbf{g}[n]$ with a specific $\tau_g$ from $\tau_g\in \{0, 0.01, 0.04, 0.1$, $0.5\}$. Color bars in these figures show the logarithm of $P_{se}$. For example, the blue color in this color bar shows the number $-3$ which represents the $P_{se}=10^{-3}$. Thus, the blue area in each subfigure is the desirable area since it offers a small error rate. 
 In Fig. \ref{tv_3d_k4}, for $K=4$, the blue area is positioned in a higher $\gamma_{t,s}$ compared to Fig. \ref{tv_3d_k16}, with $K=16$. Interestingly, for the TI channel where $\tau_g=0$, the transmit SNR by the JN has no impact on the performance of the EV-CAJ method, as shown in subfigure (a). However, by increasing $\tau_g$ and the speed of variation in $\textbf{g}[n]$,  we can see that areas in the top right of the sub-figures gradually turn into red. This implies that if the JN's signal is very strong, the channel variation is further damaging. In contrast, a signal transmitted by the JN with a weaker SNR  can be removed, even under a fast-fading channel.  This is why the blue area for $\gamma_{t,j}\le 20$ dB is quite constant for $\tau_g\le 0.04$. We can conclude that the EV-CAJ method can also effectively combat JNs with TV channels depending on the power of the JN's signal and its coherence time. Based on these results, we face a trade-off in the selection of the number of pilots and data symbols. If the JN's signal is very strong, and $\tau_g$ is large, the EV-CAJ can handle the jamming attack effectively by decreasing $t_d$ and transmitting pilot blocks more frequently. This approach reduces $\tau_g$, which leads to a significantly smaller $P_{se}$ under a strong jamming signal at the expense of a lower data rate. On the other hand, if  $\gamma_{t,j}$ and $\tau_g$ are small  (a similar scenario is depicted at the bottom of subfigures (a) and (b)), the probability of error remains still low even with fewer numbers of pilot symbols, allowing us to benefit from the increased $t_d$ as $t_p$ becomes smaller. Accordingly, we observe that for $\tau_g < 0.01$, EV-CAJ performs effectively even in the presence of strong jamming signals. Therefore, a general guideline for selecting the duration of the data block could be $t_d < 0.01 t_{c,g}$. If the jamming signal is moderate ($\gamma_{t,j} < 15$ dB), this constraint can be further relaxed to $t_d < 0.04 t_{c,g}$, allowing for a longer data block.
\subsection{Computational Complexity}
The proposed EV-CAJ and NLS-CAJ algorithms introduce computational overhead compared to scenarios where no anti-jamming (No AJ) strategy is applied. In this section, we analyze the computational complexity of both methods relative to the No AJ case. We focus on the number of required multiplications (NRM), which serves as our primary complexity metric. Multiplication of a scalar to a matrix and operations such as additions, normalizations, and selecting columns of a matrix are excluded from the analysis. Our evaluation centers on the dominant terms, which exhibit the highest polynomial order with respect to the parameters  $K$, $K_j$, $N_{t_p}$, $N_{t_s}$, and $N_{t_f}$.  To find the SVD of a matrix of size $ M \times N $, the computational complexity is given by $ O(\min(M,N)MN) $ \cite{golub1996matrix}. Therefore, we consider its complexity to be proportional to $\min(M,N)MN$ NRM for our analysis. Based on Algorithm 1, the dominant term of the  NRM for the EV-CAJ algorithm is expressed as
\begin{equation}
 R_{EV}= K N_{t_p}^2 + K^2(2N_{t_p} + N_{t_d}) - KK_j N_{t_f}.
\end{equation}
 In contrast, the NLS method utilizes $\textbf{s}_{t_p}^\perp$
  to eliminate the friendly signal from the TN. It was also developed for attacks by only one JN ($K_j=1$). Consequently, for the NLS-CAJ algorithm, the NRM can be expressed as
 \begin{equation}
 R_{NLS}= K^2(2N_{t_p} + N_{t_d}) - KK_j N_{t_f}.
\end{equation}
However, for the  No AJ case, the polynomial order of complexity is 2, whereas the NLS-CAJ and EV-CAJ algorithms have a complexity order of 3.
 This is because the No AJ case only requires the multiplication of a vector with the received matrix for channel estimation and symbol detection. Thus, the NRMs for No AJ is  $KN_{t_f}$. 
We observe that the proposed  AJ methods result in the NRM being one order of degree higher for improved performance. Specifically, if  $K_j = 1$,   $K = 4 $,  $N_{t_p} = 20$, $ N_{t_d} = 200$, and $ N_{t_f} = 220$, we find that
$R_{\text{EV}} = 4560$ and $R_{\text{NLS}} = 2960$.
In comparison, the  NRM for the No AJ case is  $880$. Thus, we note that utilizing these AJ methods necessitates a greater computational complexity.

\section{Conclusion}\label{sec_con}
In this study, a wireless network consisting of several sensing nodes was examined in the presence of a jamming attack. In this scenario, the sensing nodes cooperate with a fusion center to combat the jamming attack more effectively. Taking advantage of cooperative wireless networks, a CAJ method was proposed to combat the deceptive jamming attack. To achieve this, two methods for estimating the jamming channel direction were proposed where the magnitude of the channel vector is not necessary to be estimated. Using the proposed  EV and NLS estimation approaches, analytical formulas for the outage probability of the proposed CAJ method were derived, which showed a high agreement with simulations. Moreover, extensive simulations showed that, with a sufficient number of pilot symbols,  the EV approach can practically reach the performance of the perfect CSI case. With a sufficient number of SNs, the EV-CAJ method performed excellently in removing jamming, requiring only a $0.7$ dB higher SNR to achieve the same probability of error as the case where no jamming attack exists. The proposed method was also extended to handle multiple jamming nodes if enough DoF were available. The scenario of fast-fading channels and the required interval between pilot blocks was also discussed. The related simulation results were utilized to assess the resilience of the proposed EV-CAJ method against fast-fading channels. In future work, we intend to integrate an RL agent into the EV-CAJ method to optimize the time frame under jamming and fast-fading conditions, and to evaluate its performance in non-ideal control channels.

\appendices
\section{MLE Derivation of the signal by multiple JNs}\label{appexA}
Here, by extending the number of JNs to $K_j$, we rewrite (\ref{eq31nn}) as $\textbf{Z}_{t_p}= \textbf{Y}_{t_p} {\textbf{S}^{\perp}}^{*} =   \textbf{A}_{t_p}+\textbf{N}_{t_p}$, where $\textbf{A}_{t_p}=\sum_{l=1}^{K_j}\textbf{g}_j\textbf{a}_{t_p,l}^{T}$ and $\textbf{N}_{t_p}=\textbf{W}_{t_p}{\textbf{S}^{\perp}}^{*}$. In addition, $\textbf{a}_{t_p,l}= {\textbf{S}^\perp}^H\textbf{j}_{t_p, l}$  and $\textbf{j}_{t_p,l}$ and $\textbf{g}_l$ are the transmitted sequence and channel of the $l$-th JN, respectively. Using properties of the trace of a matrix, the LLF of $\textbf{Z}_{t_p}$ based on (\ref{eq14n}) is 
\begin{equation}\label{eq36n}
L(\textbf{Z}_{t_p})=c_o-\sigma^{-2}\mathrm{tr}[(\textbf{Z}_{t_p}-\textbf{A}_{t_p})^H(\textbf{Z}_{t_p}-\textbf{A}_{t_p})].
\end{equation}
Given the fact that  $\mathrm{rank}(\textbf{A}_{t_p})=K_j$, the MLE of $\textbf{A}_{t_p}$ can be found by minimizing  $c(\textbf{A}_{t_p})=\mathrm{tr}[(\textbf{Z}_{t_p}-\textbf{A}_{t_p})^H(\textbf{Z}_{t_p}-\textbf{A}_{t_p})]$ subject to $\mathrm{rank}(\textbf{A}_{t_p})=K_j$. Through the properties of the trace of a matrix,  the cost function can also be written as 
\begin{equation}
   c(\textbf{A}_{t_p})= \sum_{i=1}^{K}\zeta^2_i,
\end{equation}
where $\zeta_i$ is the $i$-th largest singular value of $\textbf{Z}_{t_p}-\textbf{A}_{t_p}$, and $\zeta_1\ge, ..., \zeta_{N_{t_p}}$. Here, we also assume that $K_j<K<N_{t_p}$. Since $c(\textbf{A}_{t_p})$ contains a sum of positive terms, an estimation that removes the largest positive terms and satisfies the constraint is the MLE.
Thus, we estimate the MLE of $\textbf{A}_{t_p}$ using the  $K_j$ largest singular values of $\textbf{Z}_{t_p}$  as
\begin{equation}\label{eq37nn}
\widehat{\textbf{A}_{t_p}}=\sum_{i=1}^{K_j}\lambda_i\textbf{u}_i\textbf{v}_i^H,
\end{equation}
where $\lambda_i$, $\textbf{u}_i$, and $\textbf{v}_i$ are the ordered singular values and the left and the right eigenvectors of $\textbf{Z}_{t_p}$, respectively.
The MLE in (\ref{eq37nn}) satisfies the rank constraint $\mathrm{rank}(\textbf{A}_{t_p})=K_j$ and,  concurrently, minimizes the cost function and completes the proof of (\ref{eq32nn}).

%
\bibliographystyle{IEEEtran}
\bibliography{main}





\ifCLASSOPTIONcaptionsoff
  \newpage
\fi

\end{document}